\documentclass[aps,prb,twocolumn,groupedaddress]{revtex4}
\pdfoutput=1 

\usepackage{amsmath}
\usepackage{epsfig}         % pour inclure des figures
\usepackage{psfrag}
\usepackage{ulem}
\newcommand{\bra}[1]{\langle #1|}
\newcommand{\ket}[1]{|#1\rangle}

\usepackage{color}

\newcommand{\corr}[1]{#1}

\newcommand{\ahat}[1]{\hat{#1}}
\newcommand{\hhat}[1]{\hat{#1}}

\begin{document}

\title{Quantum theory of spontaneous and stimulated emission of surface plasmons}

\author{Alexandre Archambault}
\affiliation{Laboratoire Charles Fabry,
Institut d Optique, CNRS, Univ Paris-Sud, 
Campus Polytechnique, RD128, 91127 Palaiseau cedex}

\author{Fran\c{c}ois Marquier}
\affiliation{Laboratoire Charles Fabry,
Institut d Optique, CNRS, Univ Paris-Sud, 
Campus Polytechnique, RD128, 91127 Palaiseau cedex}
\email{francois.marquier@institutoptique.fr}

\author{Christophe Arnold}
\affiliation{Laboratoire de Photonique et de Nanostructures, CNRS, Route de Nozay, 91460 Marcoussis, France}

\author{Jean-Jacques Greffet}
\affiliation{Laboratoire Charles Fabry,
Institut d Optique, CNRS, Univ Paris-Sud, 
Campus Polytechnique, RD128, 91127 Palaiseau cedex}

\date{\today}

\begin{abstract}
We introduce a quantization scheme that can be applied to surface waves propagating along a plane interface. An important result is the derivation of the energy of the surface wave for dispersive non-lossy media without invoking any specific model for the dielectric constant. Working in Coulomb's gauge, we use a modal representation of the fields. Each mode can be associated with a quantum harmonic oscillator. We have applied the formalism to derive quantum-mechanically the spontaneous emission rate of surface plasmon by a two-level system. The result is in very good agreement with Green's tensor approach in the non-lossy case. Green's approach allows also to account for losses, so that the limitations of a quantum approach of surface plasmons are clearly defined.  Finally, the issue of stimulated versus spontaneous emission has been addressed. Because of the increasing density of states near the asymptote of the dispersion relation, it is quantitatively shown that the stimulated emission probability is too small to obtain gain in this frequency region.
\end{abstract}


\maketitle

\section{Introduction}
Quantum theory of light is a useful tool to describe microscopic interactions between light and matter. The electromagnetic state is represented by photon number states and the electromagnetic field becomes an operator\cite{Loudon}. Such a description of light provides a quantitative description of absorption, spontaneous and stimulated emission of photons by a two-level system.
In particular, it allows to derive a quantitative treatment of light amplification.
It also predicts pure quantum effects, such as photon coalescence
or antibunching.
Quantum theory of light can be extended to non-dispersive  and non-lossy media.
Each photon in the material corresponds to
the excitation of
a mode characterized by a wave vector $\mathbf{k}$ and circular frequency $\omega$, such as $k= n \omega/c$, where $n$ is the refractive index of the medium and $c$ the light velocity in a vacuum. It is the purpose of this paper to introduce a quantification scheme for surface waves propagating along an interface.

It is well known that electromagnetic surface waves called surface plasmons exist at interfaces between metals and dielectrics\cite{Raether}.
Their quantum nature has been demonstrated by energy loss spectroscopy experiments on thin metallic films reported by Powell and Swan\cite{Powell}.
Single optical plasmons have been excited recently along a metallic nanowire\cite{Lukin,Wrachtrup}
Surface plasmons are associated with collective oscillation of free electrons in the metal at the surface. Similar electromagnetic fields exist also on polar materials and are called surface phonon-polariton. Both surface plasmon-polaritons and surface phonon-polaritons propagate along the interface and decrease in the direction perpendicular to the surface. Such a resonance is therefore called surface wave in a more general way. Most studies deal with a plane interface between air or vacuum and a non-lossy material. In this case, it is well known\cite{Raether} that a surface wave can exist if the dielectric constant $\epsilon(\omega)$ has a real part lower than $-1$.

Losses are often a serious limitation for many practical applications envisionned for surface plasmons. This problem could be circumvented by introducing gain in the system. Studies have been made in such a way with metallic nanoparticles embedded in a gain medium both numerically with dye molecules \cite{Smuk2006125} or quantum dots \cite{Bergman2003027402, Li2005115409} and experimentally \cite{Noginov2007455}. Seidel et al. reported the first experiment demonstrating the amplification of surface plasmons on a flat silver film surrounded by a solution of dye molecules \cite{Seidel2005177401}.
Since then, a few studies have dealt with stimulated emission of surface plasmons on flat interfaces both experimentally \cite{Ambati,NoginovPRL} and theoretically \cite{Berini}.
Such works have paved the way to active plasmonics \cite{Oulton,Stockman,Brongersma}
and nanolasers \cite{Protsenko2005063812, Bergman2004409}
or more precisely to spasers \cite{Bergman2003027402,ZheludevLasingSpaser},
or surface plasmon amplification by stimulated emission introduced by Bergman and Stockman and demonstrated experimentally recently\cite{NoginovLaser,OultonLaser}.

It is clear that a quantum treatment of surface plasmon could be useful for many applications. For instance an efficient single photon emitter could be optimized\cite{Barnes2002}. A quantum treatment allows to model stimulated emission and therefore to specify gain conditions and laser operation. It could also allow to analyse pure quantum effects for surface plasmons such as single plasmon interferences, quantum correlations\cite{fasel}, bunching, strong coupling regime\cite{Lukin2,Vuckovic} or single photon excitation of surface plasmon\cite{Altewischer, Moreno, Fasel, Zayats}. To our knowledge, the first quantization scheme for surface plasmon on a metallic surface has been reported by Elson and Ritchie \cite{Elson19714129}.
In
%this
their
work, the metal is characterized by  a non-lossy Drude model
so that real optical properties cannot be included.
Using Green's approach, Gruner and Welsch introduce a quantization scheme for electromagnetic fields in dispersive and absorptive materials \cite{Gruner19961818}. It should hence be possible to quantize the field associated with surface waves using their model. Note that due to losses, they cannot obtain operators for modes but only local operators : one recovers the usual creation/annihilation operators in the limit of zero losses. A related work, reported in the early nineties by Babiker et al., dealt with the quantization of interface optical phonons in quantum well, which could appear also as a confined surface phonon in a heterostructure\cite{Babiker19932236}.

In this paper, a quantization scheme that is not based on a specific model of the dielectric constant is introduced. The aim is to quantize the field by accounting for the experimental dispersion properties of the medium. The procedure follows the quantization scheme for photons in a vacuum. We will first introduce a classical mode description of the surface waves and discuss the dispersion relation. A key issue for quantization is the definition of the energy of surface waves for dispersive lossy media. The problem of electromagnetic energy in a dispersive and lossy medium has been recently addressed in a paper by Stallinga \cite{Stallinga2006026606}. The third section addresses the problem of the electromagnetic energy associated with surface waves in a simpler case following Landau and Lifchitz for non-lossy dielectric material \cite{Landau}. The quantization scheme is finally described in the fourth section.
In order to check our results, we apply our formalism in the fifth section to the calculation of the spontaneous emission of a two-level system in
the
presence of surface plasmons.
The Purcell factor (i.e. the local density of states normalized by the vacuum density of states)
and Einstein's coefficients are also derived using this model.

\section{Modal description of surface waves}
\begin{figure}[htbp]
\includegraphics[width=8cm]{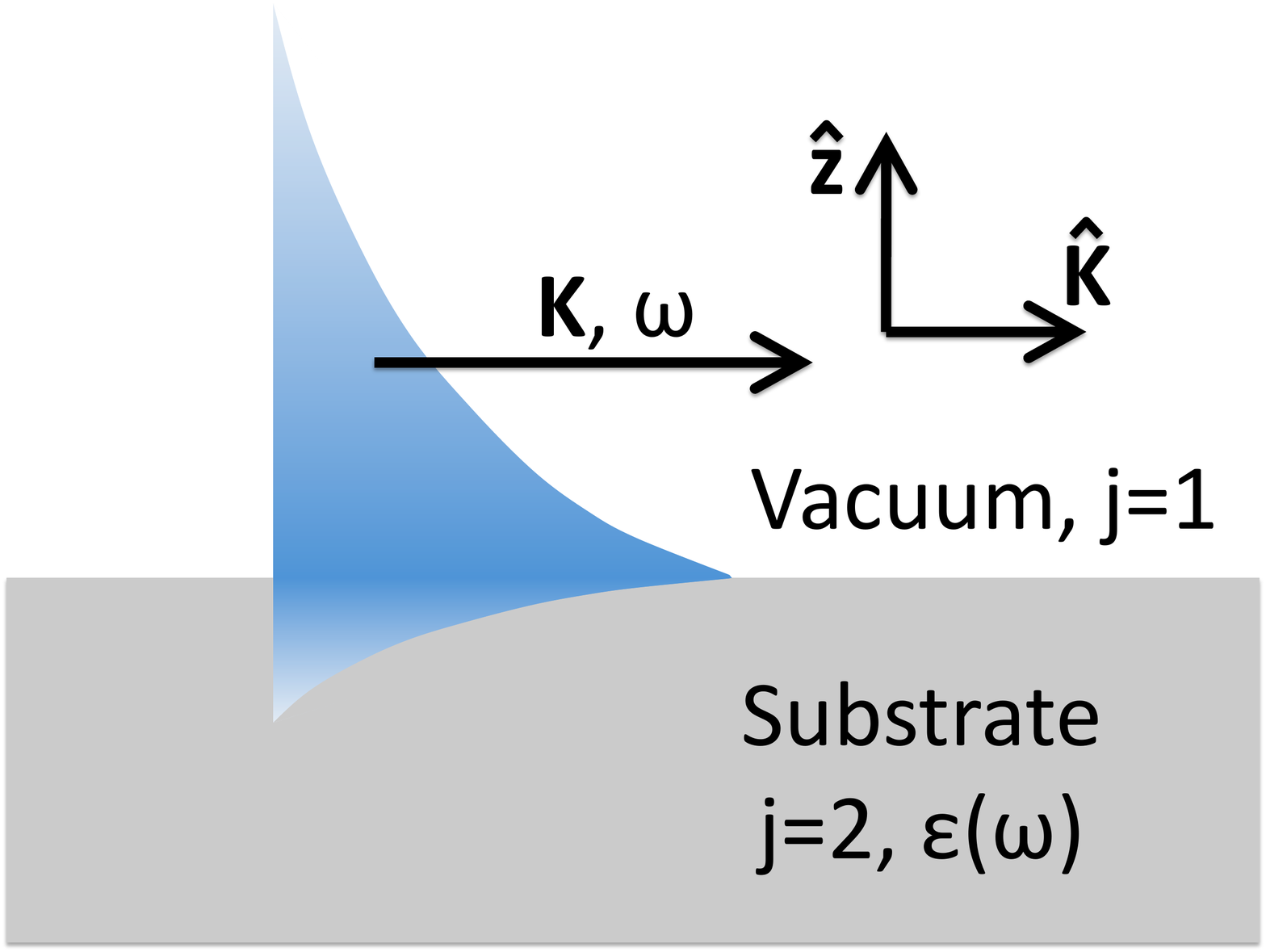}
\caption{Surface wave on a plane interface. The surface mode is characterized by its circular frequency $\omega$ and the projection of the wave vector along the interface $\mathbf{K}$. $\mathbf{\hat{K}}$ and $\mathbf{\hat{z}}$ are unit vectors along and perpendicular to the plane interface respectively. \label{config}}
\end{figure}

Let us consider surface waves propagating on a plane interface at $z=0$ separating two semi-infinite media (Fig.~\ref{config}). One of them is a vacuum or air and the second is a metal or a polar material. A surface mode is characterized by its circular frequency $\omega$ and the projection of the wave vector  $\mathbf{K}$ on the plane perpendicular to the $z$-axis. The material has a dielectric constant $\epsilon(\omega)$. We use Coulomb's gauge ($\mathrm{div}\mathbf{A}(\mathbf{r},t) = 0$) to write the magnetic and electric fields:
\begin{eqnarray}
\mathbf{B}(\mathbf{r},t) =\mathbf{\nabla\times}\mathbf{A}(\mathbf{r},t)\label{BA} \\
\mathbf{E}(\mathbf{r},t) =-\frac{\partial\mathbf{A}(\mathbf{r},t) }{\partial t}\label{EA}
\end{eqnarray}

The field produced by any distribution of sources in the presence of an interface can be computed using Green's tensor. By extracting the pole contribution, it is possible to derive the general form of the surface plasmon field. The details of this procedure can be found in Ref.~\onlinecite{PRBAlex}.
The corresponding vector potential can be cast in the form:
\begin{equation}
\mathbf{A} (\mathbf{r},t)  =
	\int \frac{\mathrm{d}^2 \mathbf{K}}{(2\pi)^2} \ 
 \alpha_{\mathbf{K}} \mathbf{u} _{\mathbf{K}} (z) \  \exp(i\mathbf{K}.\mathbf{r}) \exp(-i\omega_{sp} t) + c.c.
\end{equation}
where $c.c.$ stands for complex conjugate.
In this equation, $\mathbf{K}$ is a real wave vector parallel to the interface and
the circular frequency $\omega_{sp}$ is a complex root of the equation:
\begin{equation}\label{reldisp}
K=\frac{\omega}{c}\sqrt{\frac{\epsilon(\omega)}{\epsilon(\omega)+1}}
\end{equation}
The term $\alpha_{\mathbf{K}}$ is an amplitude associated with wave vector $\mathbf{K}$ in the decomposition.
The vectors $\mathbf{u} _{\mathbf{K}} (z)$ are given by:
\begin{eqnarray}
\label{Uj}
\mathbf{u} _{\mathbf{K}} (z) &= \dfrac{1}{\sqrt{L(\omega_{sp})}} \exp(i\gamma _j  z )
		\left(\mathbf{\hat{K}} - \dfrac{K}{\gamma _j} \mathbf{\hat{z}} \right)
\end{eqnarray}
where $L(\omega_{sp})$ has the dimension of a length and will be fixed later by Eq.~\eqref{NormalizationL} to normalize the energy of each mode.
$\gamma_j$ is the projection of the wave vector along the $z$-axis,
$j=1$ in the region $z>0$, and $j=2$ in the region $z<0$,
so that
$\gamma_j^2=\epsilon_j(\omega_{sp})\omega_{sp}^2/c^2-K^2$.
The sign of $\gamma_j$ is then chosen such as
the field goes to zero when $z$ goes to $\pm\infty$.
Let us note that in the non-lossy case, $\gamma_1$ and $\gamma_2$ are purely imaginary, so that the electric field decays exponentially along the $z$-axis.
$\mathbf{\hat{K}}$ and $\mathbf{\hat{z}}$ are unit vectors directed along $\mathbf{K}$ and the $z$-axis respectively.

\begin{figure}[htbp]
\includegraphics[width=8cm]{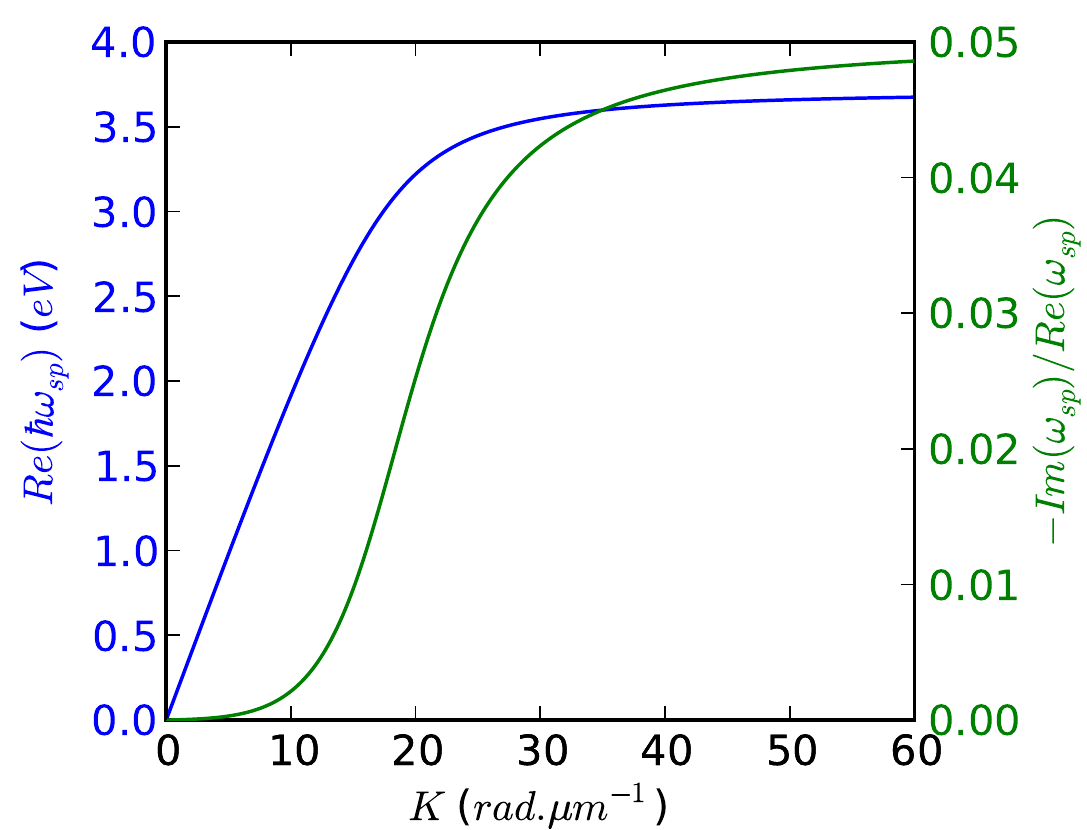}
\caption{Dispersion relation of a surface plasmon on a plane interface between air and silver (solid line, left axis) and variation of the imaginary part of $\omega$ (dashed line, right axis). The dispersion relation has been obtained using  the silver dielectric constant given in Appendix~\ref{modeleps} \label{reldispfig}}
\end{figure}

Figure~\ref{reldispfig} shows the dispersion relation of surface plasmons as well as the variation of the imaginary part of the frequency with $K$ on a plane interface of silver. 
To perform the calculations when a complex frequency is needed, it has been useful to fit the experimental values of the dielectric constant $\epsilon(\omega)$ given by Ref.~\onlinecite{Palik} with an analytical model. The real part of the silver dielectric constant is very well represented by a Drude model given in Ref.~\onlinecite{Oulton}. For the imaginary part we add to this Drude model a conductivity term, so that the modelized dielectric constant is in very good agreement with the experimental data. The model we used is given in appendix~\ref{modeleps}.
In this example, Im($\omega_{sp}$) is small, less than $5\%$, comparing to Re($\omega_{sp}$). 
In other words, the lifetime of the surface mode is long enough to have a few tens of periods for the oscillating electromagnetic field on the asymptotic part of the dispersion relation and hundreds of periods on the linear part, close to the light cone. Note that this point seems to be rather general. Indeed we found similar ratios for many other materials supporting surface waves.

From the dispersion relation, it is possible to derive the density of states. To this aim, it is convenient to introduce a virtual box, which is in fact a virtual square in the $x-y$ plane of sides $L_x$ and $L_y$ and size $S=L_x \times L_y$.
Born-Von Karman's conditions yields a quantized value of the wave vector $K_x=n_x2\pi/L_x, K_y=n_y2\pi/L_y$, where $n_{x,y}$ are relative integers. Let us note that a different expansion of the surface plasmon field can be used with a complex wave vector and a real frequency as discussed in Ref.~\onlinecite{PRBAlex}. We stress here that the Born-von Karman procedure imposes a real wave vector. It follows that the relevant dispersion relation has no backbending as seen in Fig.~\ref{reldispfig}. The reader is refered to Ref.~\onlinecite{PRBAlex} for more details.
Substituting the discrete sum $\frac{1}{S}\sum_\mathbf{K}$ over the quantized wave vector $\mathbf{K}$ and
the discrete amplitude $S A_\mathbf{K}$ to $\int \frac{\mathrm{d}^2 \mathbf{K}}{(2\pi)^2}$ and $\alpha_\mathbf{K}$ respectively,
the vector potential can be  cast as
\begin{equation}
	\label{expA}
	\mathbf{A}(\mathbf{r},t) =
		\sum_{\mathbf{K}} A_{\mathbf{K}} \mathbf{u} _{\mathbf{K}} (z)
					\exp(i \mathbf{K}. \mathbf{r}) \exp(-i\omega t)
		+ c.c.
\end{equation}
where we have omitted the subscript $sp$ for the circular frequency $\omega$.
We can insert this form in  Eqs.~(\ref{BA}) and (\ref{EA}) to obtain the electric and magnetic fields.
Introducing the notations $\mathbf{k}_j = \mathbf{K} + \gamma_j\mathbf{\hat{z}}$ and
$\mathbf{b}_{\mathbf{K}} (z) = \mathbf{k}_j \times \mathbf{u} _{\mathbf{K}} (z)$, we have:
\begin{subequations}
\label{expEB}
\begin{eqnarray}
	\mathbf{E}(\mathbf{r},t) &=&
		i \sum_{\mathbf{K}} \omega A_{\mathbf{K}} \mathbf{u} _{\mathbf{K}} (z)
					\exp(i \mathbf{K}. \mathbf{r}) \exp(-i\omega t)
		+ c.c.
	\label{expE}
	\\
	\mathbf{B}(\mathbf{r},t) &=&
		i \sum_{\mathbf{K}} A_{\mathbf{K}} \mathbf{b} _{\mathbf{K}} (z)
					\exp(i \mathbf{K}. \mathbf{r}) \exp(-i\omega t)
		+ c.c.
	\label{expB}
\end{eqnarray}
\end{subequations}

\section{Energy of a surface wave}
\label{secenergy}

The quantization procedure is based on the fact that the energy of the field has the structure of a sum of harmonic oscillators. It is thus a key issue to derive the energy of the surface plasmon field. 
In this section, we give a brief outline of the derivation and leave the details to appendix~\ref{energy}.  In a vacuum, the energy density is given by\cite{Jackson}:
\begin{equation}
u_{1} = \frac{\epsilon_0}{2}\mathbf{E}^2(\mathbf{r},t)+\frac{1}{2\mu_0}\mathbf{B}^2(\mathbf{r},t)
\end{equation}\\

The electromagnetic energy in a lossy dispersive material is a more subtle issue. This problem has been addressed for the first time by Brillouin\cite{Brillouin}.
He considered a very simple case, with two perfectly monochromatic waves in the material.
Landau and Lifchitz analysed\cite{Landau} the energy of an electromagnetic field in a non-lossy dispersive medium, whose frequencies form a narrow continuum around the mean frequency $\omega_0$. They dealt with fields such as $\mathbf{E}=\mathbf{E}_0(t)\exp(-i\omega_0 t)$, $\mathbf{E}_0(t)$ varying slowly over the period $2\pi/\omega_0$. 
In the appendix, we follow this method. The main idea is to derive the work done by an external operator to build adiabatically the field amplitude. This work is equal to the total amount of electromagnetic energy of the surface waves for a non-lossy medium. Note that more recently, Stallinga derived an expression of the energy for dispersive and lossy materials\cite{Stallinga2006026606}. The result is the same provided that $\epsilon$ is replaced by Re$(\epsilon)$.
This suggests that it is possible to neglect losses in the calculation of the energy. Actually, it is essential to deal with a non-lossy medium to have well-defined modes.
A key issue regarding this approximation is whether the dispersion relation is modified by the presence of losses.
Indeed, the density of states critically depends on the dispersion relation.
We compared the dispersion relation obtained using Re$(\omega_{sp})$ for a lossy medium with the dispersion relation with a non-lossy medium in the case of silver.
We found a relative difference between the two dispersion relations always less than
$1.5\times10^{-3}$.

We will thus
neglect the losses of the medium in the derivation of the energy. The calculation outlined in appendix~\ref{energy} gives the total energy of the surface waves:
\begin{equation}\label{exprenergie}
U = \sum_{\mathbf{K}}  \epsilon_0 \omega ^2 S  \left[ A_{\mathbf{K}} A_{\mathbf{K}}^* + A_{\mathbf{K}}^* A_{\mathbf{K}} \right].
\end{equation}
We emphasize that this convenient expression for the energy is obtained using
the right normalization condition on
$L(\omega)$
or equivalently on
$\mathbf{u}_{\mathbf{K}} (z)$
given respectively
by Eqs.~\eqref{NormalizationL} and \eqref{NormalizationU}.

\section{Quantization of surface waves}
\label{sec_quantization}

We now turn to the quantization of the electromagnetic field of surface plasmons. We first notice that the expression
$\epsilon_0 \omega ^2  S \left[ A_{\mathbf{K}} A_{\mathbf{K}}^* + A_{\mathbf{K}}^* A_{\mathbf{K}} \right]$
of the energy for each mode $\mathbf{K}$, has the structure of the energy of a harmonic oscillator,
hence the quantized hamiltonian:
\begin{equation}
\hhat{H} = \sum _{\mathbf{K}} \frac{\hbar\omega}{2}\left[\ahat{a}_{\mathbf{K}} \ahat{a}_{\mathbf{K}}^{\dag} + \ahat{a}_{\mathbf{K}}^{\dag} \ahat{a}_{\mathbf{K}}\right]
\end{equation}
with the equivalence
\begin{eqnarray}
A_{\mathbf{K}} & \rightarrow  & \sqrt{\frac{\hbar}{2\epsilon_0 \omega S}} \,\ahat{a}_{\mathbf{K}} \\
A_{\mathbf{K}}^* & \rightarrow & \sqrt{\frac{\hbar}{2\epsilon_0 \omega S}}\,\ahat{a}_{\mathbf{K}}^{\dag}.
\end{eqnarray}

The surface wave field is thus quantized by association of a quantum-mechanical harmonic oscillator to each mode $\mathbf{K}$.
We introduce $\ahat{a}_{\mathbf{K}}^{\dag}$ and $\ahat{a}_{\mathbf{K}}$ which are respectively the creation and annihilation operators for the
mode $\mathbf{K}$.
As in the harmonic oscillator theory, $\ahat{a}_{\mathbf{K}}^{\dag}$ and $\ahat{a}_{\mathbf{K}}$ act on surface wave number states $\ket{n_{\mathbf{K}}}$ which are eigenvectors associated with eigenvalues $(n_{\mathbf{K}}+1/2)\hbar\omega$ of the Hamiltonian ($n_{\mathbf{K}}$ is an integer). Operators $\ahat{a}_{\mathbf{K}}^{\dag}$ (respectively  $\ahat{a}_{\mathbf{K}}$) allow to create (respectively destroy) a quantum of energy $\hbar\omega$ according to the operating rules\cite{Loudon}:
\begin{eqnarray}
\ahat{a}_{\mathbf{K}}^{\dag} \ket{n_{\mathbf{K}}} & = & \sqrt{n_{\mathbf{K}}+1} \ket{n_{\mathbf{K}}+1} \label{a+}\\ 
\ahat{a}_{\mathbf{K}} \ket{n_{\mathbf{K}}} & = & \sqrt{n_{\mathbf{K}}} \ket{n_{\mathbf{K}}-1} \label{a}
\end{eqnarray}

Different surface modes are independent so that their associated operators commute:
\begin{equation}
[\ahat{a}_{\mathbf{K}} ,\ahat{a}_{\mathbf{K}'}^{\dag} ]=\delta_{\mathbf{K},\mathbf{K}'}.
\end{equation}

We can now write the fields  as operators acting on the surface plasmon number quantum states  $\ket{n_{\mathbf{K}}} $:

\begin{multline}\label{Equantif}
\mathbf{\hat{E}} (\mathbf{r},t) =
	i \sum_{\mathbf{K}} \ 
\sqrt{\frac{\hbar\omega}{2\epsilon_0 S }}
\mathbf{u} _{\mathbf{K}} (z) \   \ahat{a}_{\mathbf{K}} \exp(i\mathbf{K}.\mathbf{r}) \exp(-i\omega t) \\
 + h.c.
\end{multline}
\begin{multline}\label{Bquantif}
\mathbf{\hat{B}}(\mathbf{r},t) =
	i \sum_{\mathbf{K}} \ 
  \sqrt{\frac{\hbar}{2\epsilon_0 \omega S}} \mathbf{b} _{\mathbf{K}} (z) \  \ahat{a}_{\mathbf{K}} \exp(i\mathbf{K}.\mathbf{r}) \exp(-i\omega t) \\
+ h.c.
\end{multline}
where $h.c.$ denotes the hermitian conjugate.

\section{Emission rates: comparison with the classical case, Einstein's coefficients}

\subsection{Spontaneous emission of a dipole above a metallic interface}\label{dipole}

The quantization scheme that we have introduced allows to derive an expression of the electromagnetic field using operators.
Hence, we can write interaction hamiltonians and describe the coupling between light and matter. In order to test this quantization procedure, we performed the calculation of the lifetime of a two-level system placed in the vicinity of a metal-vacuum interface so that surface plasmons can be excited. This result is interesting as the lifetime can also be computed using a classical approach as shown for instance by Ford and Weber~\cite{Ford1984195}. More specifically, they showed how to find the surface plasmon contribution to the lifetime by extracting the pole contribution. By comparing both results, we can assess the validity of the quantum theory of surface plasmon within the approximation of a dispersive but non-lossy medium. 

\subsubsection{Quantum calculation}
\label{secquantumcalc}
In the quantum approach, we first derive the decay rate associated to the spontaneous emission of surface plasmons
of a two-level quantum system close to an interface, using Fermi's golden rule.
This gives the surface plasmon spontaneous emission rate as a function of the
matrix element $\langle 2 \vert \mathbf{\hat{D}} \vert 1 \rangle = \mathbf{D}_{12}$ of the dipole moment operator
$\mathbf{\hat{D}}$.
The details of the calculation are given in appendix \ref{quantumcalc}.

We obtain the following expression for the spontaneous emission rate:
\begin{multline}
	\label{Rate}
	\gamma_{spont} (\mathbf{D}_{12}, \omega_0, z) =
		\frac{\omega_0 \lvert \mathbf{D_{12}} \rvert ^2}{2 \epsilon_0 \hbar}
			K\frac{\mathrm{d} K}{\mathrm{d} \omega}
			\frac{1}{L_{eff} (z,\mathbf{d}_{12},\omega_0)}
\end{multline}
in which
$\mathbf{d}_{12} = \mathbf{D}_{12} / \lvert \mathbf{D}_{12} \rvert$ is
the (possibly complex) polarization of the dipole,
$d_{12,z} = \mathbf{d}_{12}.\mathbf{\hat{z}}$,
$\mathbf{d}_{12,/\!/} = \mathbf{d}_{12} - d_{12,z} \mathbf{\hat{z}}$.
We introduced the effective length of the surface plasmon mode
$L_{eff} (z,\mathbf{d}_{12},\omega_0)$,
\begin{multline}
\label{Leff}
\frac{1}{L_{eff} (z,\mathbf{d}_{12},\omega_0)}
			= \frac{\exp(2i\gamma_1 z)}{L(\omega_0)}
		\left[ \frac{1}{2} \lvert \mathbf{d}_{12, /\!/} \rvert ^2 - \epsilon(\omega_0) \lvert d_{12, z} \rvert ^2 \right].
\end{multline}
It will be seen later that this length allows to define an effective volume of the plasmon mode.

For comparison with the classical calculation, we normalize $\gamma_{spont} (\mathbf{D}_{12}, \omega_0, z)$
with the spontaneous
emission rate of the same two-state quantum system in a vacuum, given by\cite{Loudon} $\gamma_{spont} ^0 = \frac{\omega_0 ^3 \lvert \mathbf{D}_{12} \rvert ^2}{3\pi \epsilon_0 \hbar c^3}$.
This gives the Purcell factor associated to the emission of surface plasmons:
\begin{multline}
	\label{RateNormalized}
	F_P (\mathbf{d}_{12}, \omega_0, z) =
		\frac{3\pi c^3}{2 \omega_0 ^2}
			K\frac{\mathrm{d} K}{\mathrm{d} \omega}
			\frac{1}{L_{eff} (z,\mathbf{d}_{12},\omega_0)}
\end{multline}
which does not depend anymore on the amplitude of $\mathbf{D}_{12}$, but only on its polarization $\mathbf{d}_{12}$,
its frequency $\omega_0$ and its distance to the interface $z$. As expected, the Purcell factor
decreases exponentially as the dipole goes farther from the interface, and can have
rather high values (see Fig.~\ref{lifetime2} and comments below) as $\omega _0$ gets closer
to the asymptotic frequency of surface plasmons if the dipole is not too far from the interface.

This Purcell factor can also be cast under the form:
\begin{multline}
\label{PurcellF}
	F_P (\mathbf{d}_{12}, \omega_0, z) =
		\omega_0 \ g(\omega_0)   \ 
			\frac{\lambda_0 ^3}{V_{eff}(z,\mathbf{d}_{12},\omega_0)} \ 
		\frac{3}{8\pi}
\end{multline}
where the (global) density of states of surface plasmons $g (\omega)$ is given by
$g (\omega) = S \frac{K}{2\pi} \frac{\mathrm{d}K }{\mathrm{d} \omega}$
and
$V_{eff}(z,\mathbf{d}_{12},\omega_0) = S L_{eff}(z,\mathbf{d}_{12},\omega_0)
 $
is the volume of the surface plasmon modes of frequency
$\omega_0$ for a dipole polarization $\mathbf{d}_{12}$ in which the emission occurs.
Eq.~\eqref{PurcellF} is thus
similar to the Purcell factor $F_P$ of a dipole interacting
with a single damped mode\cite{Purcell}.
($F_P = Q \frac{\lambda^3}{V} \frac{3}{4\pi ^2}$,
or equivalently
$F_P = \omega g (\omega) \frac{\lambda^3}{V} \frac{3}{8\pi}$
using the density of states of the single mode at resonance,
$g(\omega) = \frac{2}{\pi} \frac{Q}{\omega}$.)

When dealing with an isotropic distribution of dipoles, the average of
the rate of spontaneous emission~\eqref{Rate}, over
the orientations of the dipole $\mathbf{D}_{12}$, should be considered.
Let us first introduce the total
effective length of the surface plasmon mode,
defined as the inverse of the average of $\frac{1}{L_{eff} (z,\mathbf{d}_{12},\omega_0)}$
over the directions of $\mathbf{d}_{12}$:
\begin{equation}
 \frac{1}{L_{eff,total} (z,\omega_0)} = 
		 \frac{1}{3}
			\frac{\exp(2i\gamma_1 z)}{L(\omega_0)}
		\left[ 1 + \lvert \epsilon(\omega_0) \rvert \right],
\end{equation}
Calculating the averaged rate of spontaneous emission then amounts
to replacing $L_{eff} (z,\mathbf{d}_{12},\omega_0)$
by $L_{eff,total} (z,\omega_0)$ in Eq.~\eqref{Rate}:
\begin{multline}
	\label{RateTotal}
	\gamma_{spont,total} (\lvert \mathbf{D}_{12} \rvert, \omega_0, z) =
		\frac{\omega_0 \lvert \mathbf{D_{12}} \rvert ^2}{2 \epsilon_0 \hbar}
			K\frac{\mathrm{d} K}{\mathrm{d} \omega} \\
			\frac{1}{L_{eff,total} (z,\omega_0)}.
\end{multline}
More details are given in appendix \ref{quantumcalc}.

\subsubsection{Classical approach}

In the previous section, we considered a two-level quantum system having
a given polarization $\mathbf{d}_{12}$ and Bohr circular frequency $\omega_0$,
and we normalized its spontaneous emission rate by its value in a vacuum.
The power radiated by a classical harmonic dipole
having the same polarization $\mathbf{d}_{12}$ and a circular frequency $\omega_0$
can also be normalized by its value in a vacuum.
\corr{Both expressions give the \corr{normalized} local density of states, which is a classical quantity. They are therefore equal,}
that is the normalized radiated power gives the normalized spontaneous emission rate.
The normalized radiated power can be expressed as a function of Green's tensor $\tensor{\mathbf{G}} (\mathbf{r}, \mathbf{r}', \omega)$
of the system:
\begin{equation}\label{Novot}
F_{P,cl} (\mathbf{d}_{12}, \omega_0, z)=
	\frac{6\pi c}{\omega_0}
	\mathrm{Im} \left[ \mathbf{d} ^\ast _{12} .
	 \tensor{\mathbf{G}}(z \mathbf{\hat{z}},z\mathbf{\hat{z}},\omega_0)\mathbf{d}_{12} \right].
\end{equation}

Following the steps detailed in Ref.~\onlinecite{PRBAlex}, the pole contribution
$\tensor{\mathbf{G}}_{sp}$
of Green's tensor of a plane interface
can be derived, and inserted in Eq.~\eqref{Novot}.
We use here the pole contribution of the surface plasmon with a complex frequency (see Ref.~\onlinecite{PRBAlex}).
The details of the calculation are given in appendix~\ref{rategreen}.
One finds for the normalized radiated power in the non-lossy case:
\begin{multline}
	\label{Gcl_ll}
	F_{P,cl} (\mathbf{d}_{12}, \omega_0, z) =
		\frac{3\pi c^3}{\omega_0 ^3}
		 K^3\frac{\mathrm{d}K}{\mathrm{d}\omega} \, 
			R(K, \omega_{0})
		\\
			\exp(2i\gamma_1 z)
	  \left[
		\frac{1}{2} \lvert \mathbf{d}_{12,/\!/} \rvert ^2
		- \epsilon(\omega _{0}) \lvert d_{12,z} \rvert ^2
	  \right].
\end{multline}
Using Eq.~\eqref{RL}
and comparing Eq.~\eqref{Gcl_ll} to
Eq.~\eqref{RateNormalized}, we see easily that $F_P (\mathbf{d}_{12}, \omega_0, z) = F_{P,cl} (\mathbf{d}_{12}, \omega_0, z)$.
We thus recover the quantum spontaneous emission rate
in the non lossy limit of the above classical approach.
This result is not surprising. \corr{Indeed the normalized spontaneous emission rate
yields the local density of states. The latter is a classical quantity.} In the quantum
approach, it has been calculated using the dispersion relation. In the classical
approach, it has been calculated using the Green's tensor.
We have thus checked that
the mode approach and the Green's formalism approach are equivalent.
We now go one step further and compare the
quantum approach (without losses) with the Green's tensor approach that
accounts for losses. We compute the spontaneous emission rate for both cases
in order to assess the role of losses.

\subsubsection{Comparison with the lossy case}
\begin{figure}[htbp]
\includegraphics[width=8cm]{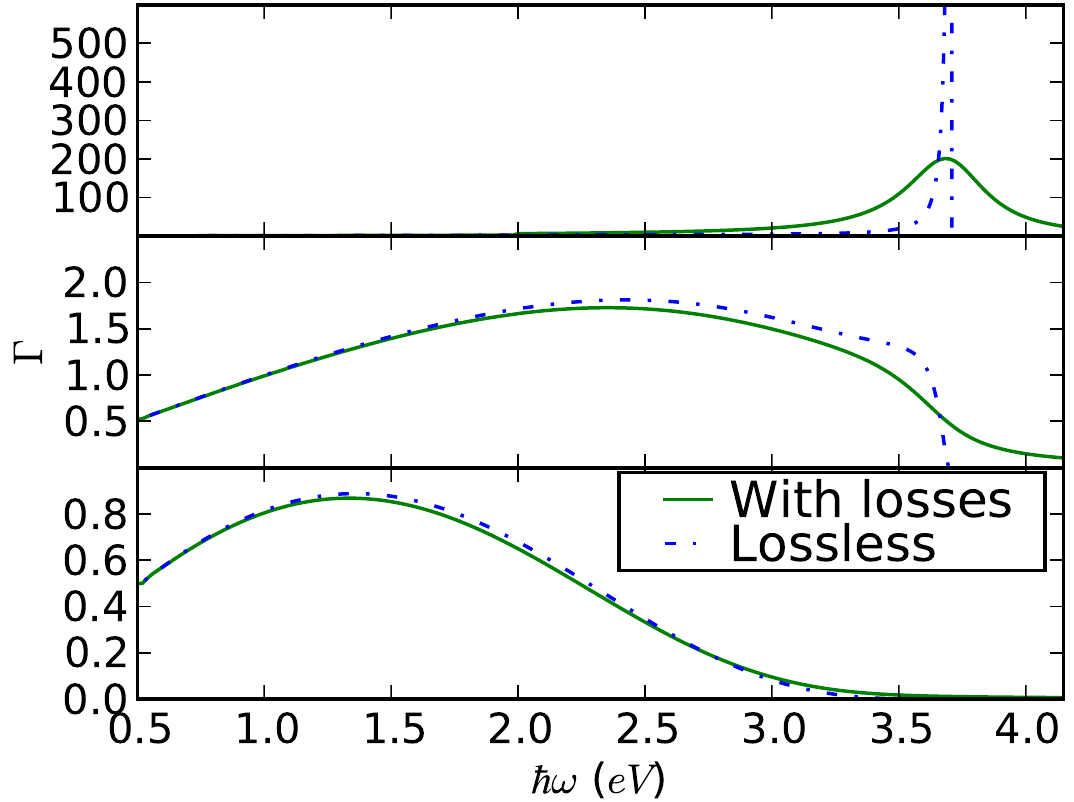}
\caption{Normalized emission rate $F_P$ of
a vertical dipole
located at
10 nm (top), 75 nm (center) and 250 nm (bottom)
from the surface when taking into account losses (green continuous curves) comparing to the non-lossy cases (blue dashdotted curves). \label{lifetime2}}
\end{figure}

The lossy and non lossy emission rates are compared using Eqs.~\eqref{RateNormalized} and \eqref{RateW}.
The result is seen on Fig.~\ref{lifetime2} for a dipole located at three different distances of a silver surface (10, 75 and 250 nm). It appears that the differences between both curves in this case are still very small as long as the frequency is not too close from the asymptote of the dispersion relation. This is not surprising considering the fact that at this asymptotic value the losses are the most important.
Moreover, when the distance between the dipole and the interface
increases,
the part of the electromagnetic field due to the higher surface plasmons wave vector decreases, so that the part due to the surface plasmons lying on the linear part of the dispersion relation is more important. As seen in Fig.\ref{reldispfig}, these surface waves have less losses and the quantum approach is thus more accurate.  
It follows the important conclusion that the non-lossy medium approximation in the quantum treatment is reasonable to deal with surface waves provided that the frequency is not too close to the asymptotic value.

Note that Ref.~\onlinecite{Ford1984195} provides an expression for
the surface plasmon emission rate of a dipole close to an interface,
which can be compared to ours  (details not given here).
In the non lossy case, it can be proved analytically
that their results lead to the same
normalized emission rate as Eq.~\eqref{RateNormalized}. In the lossy case,
one finds a normalized emission rate close to the one used here (Eq.~\eqref{RateW}),
although they are not rigorously equal. Our method gives an expression of the normalized surface
plasmon emission rate as a sum over the modes $\mathbf{K}$ (see Eq.~\eqref{RateW_sum}),
which provides a better understanding of the difference between the lossy and the non lossy cases.

\subsection{Einstein's coefficients}
A quantum approach for surface waves allows us also to derive easily Einstein's coefficients for spontaneous and stimulated emission. The same example of a dipole above the interface is taken.
Once again, it is possible to follow the approach described for photons in Ref.~\onlinecite{Loudon} for instance.
Einstein's coefficient for
surface plasmon
spontaneous emission has already been calculated: $A_{21}=\gamma_{spont}(\mathbf{D}_{12}, \omega_0, z)$ (see Eq.~\eqref{Rate}). In order to obtain Einstein's coefficient for stimulated emission, one needs to start from Eq.~(\ref{EltMat}). In this equation, the term proportional to $n_{\mathbf{K}}$ is the matrix element for stimulated emission. We note $\langle W(\omega)\rangle$ the energy density of the radiation
per unit surface
and we assume that it varies slowly for frequencies near $\omega_0$. The total energy in the single mode $n_{\mathbf{K}}$ is now replaced by:

\begin{equation}
n_{\mathbf{K}}\hbar\omega \rightarrow S \int \mathrm{d}\omega\langle W(\omega)\rangle
\end{equation}

The transition rate due to stimulated emission can thus be written:
\begin{multline}
 \gamma_{stim} (\mathbf{D}_{12}, \omega_0, z) =\\
\frac{2\pi}{\hbar^2}
	\int \mathrm{d}\omega\langle W(\omega)\rangle
	\frac{1}{2\epsilon_0}
	\left\lvert \mathbf{D}_{12} . \mathbf{u} _{\mathbf{K}} (\mathbf{r}) \right\rvert ^2 \ 
\delta(\omega-\omega_0)
\end{multline}

It follows that Einstein's coefficient for stimulated emission in mode $\mathbf{K}$,
$B_{21} = \gamma_{stim} (\mathbf{D}_{12}, \omega_0, z) / \langle W(\omega_0)\rangle$,
is given by:
\begin{equation}
B_{21} =
\frac{\pi \lvert \mathbf{D}_{12} \rvert ^2}{\epsilon_0 \hbar^2}
	\frac{\exp(2i\gamma_1 z)}{L(\omega_0)}
	\left\lvert d_{12,/\!/} \cos \phi - \frac{K}{\gamma_1} d_{12,z} \right\rvert ^2
	%\rho _{\mathbf{d}} ^{(\mathbf{K})} (z, \omega)
\end{equation}
where $d_{12,/\!/}$ and  $d_{12,z}$ are defined above, and $\phi$ is the angle
between the projection of $\mathbf{D}_{12}$ on the interface and $\mathbf{K}$.
When dealing with an isotropic distribution of dipoles, $B_{12}$ should
be averaged over the directions of $\mathbf{D}_{12}$, in the same
way as in Sec.~(\ref{secquantumcalc}). We get the total Einstein coefficient
for stimulated emission in mode $\mathbf{K}$
\begin{equation}
B_{21,total} =
\frac{\pi \lvert \mathbf{D}_{12} \rvert ^2}{3 \epsilon_0 \hbar^2}
	\frac{\exp(2i\gamma_1 z)}{L(\omega_0)}
	\left[ 1 - \epsilon(\omega_0) \right]
\end{equation}

\begin{figure}[htbp]
 \includegraphics[width=8cm]{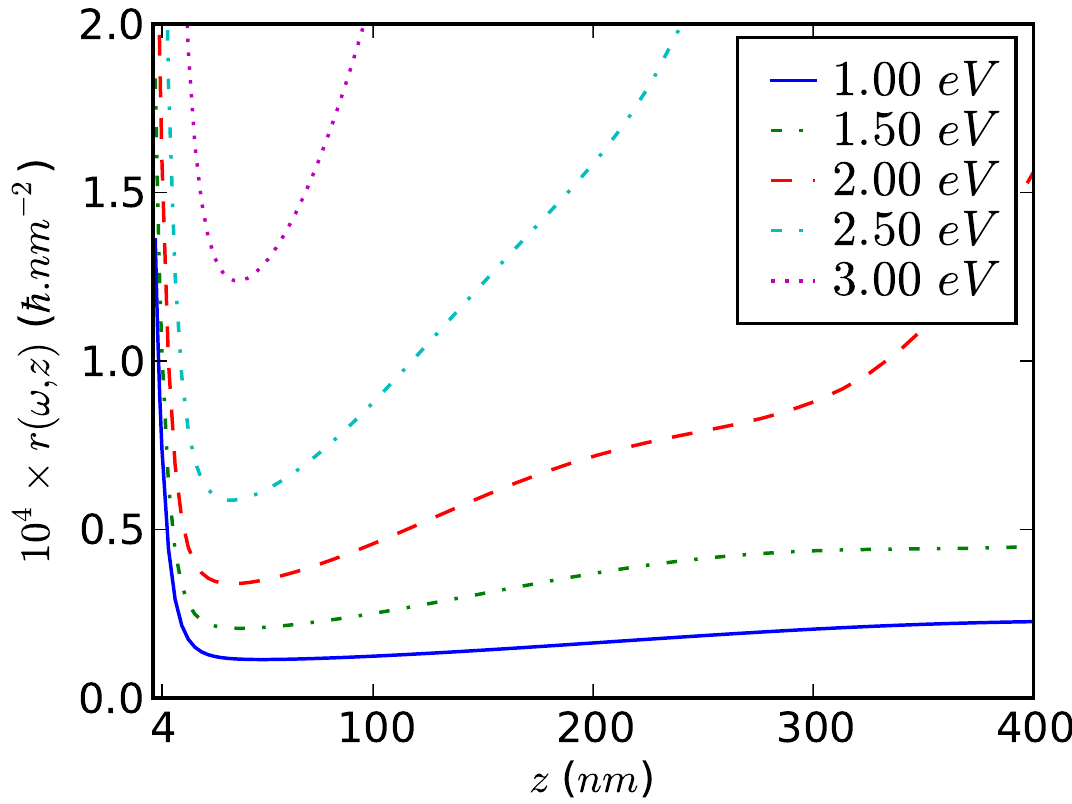}
 \caption{Ratio of Einstein's coefficients $r(\omega,z)$ for (total) spontaneous emission and
stimulated emission of surface plasmons as a function of the distance to the interface,
for several values of $\omega$ (see legend). \label{AB}}
 \end{figure}

To describe the amplification of a surface plasmon beam by an amplifying medium,
it is interesting to derive 
the ratio $r(\omega_0, z)=A_{21}^{(i)}(\lvert\mathbf{D}_{12}\rvert, \omega_0, z)/B_{21,total}(\lvert\mathbf{D}_{12}\rvert, \omega_0, z)$,
where $A_{21}^{(i)}(\lvert\mathbf{D}_{12}\rvert, \omega_0, z) =
 \gamma_{spont} ^0
\langle F ^{(i)} _{P,cl} (\mathbf{d}_{12}, \omega_0, z) \rangle$
stands for the
total spontaneous emission rate of the dipole close to the interface ($(i)$ denotes interface),
and $\gamma_{spont}^0$ is given above.
It can be computed with Eq.~\eqref{Novot},
using Green's tensor of a plane interface (this rate includes all the waves that can be emitted,
not only surface plasmons). $\langle \cdot \rangle$
stands for average over the orientations $\mathbf{d}_{12}$ of the dipole.
$r(\omega_0, z)$ gives the threshold energy per unit surface $W_c (\omega_0)$ at which
the stimulated emission rate equals the spontaneous  one.
It 
can be written as:
\begin{equation}
	r(\omega_0, z) =
		r^0(\omega_0)
		\frac{\langle F_{P,cl} ^{(i)} (\mathbf{d}_{12}, \omega_0, z) \rangle}{\exp(2i\gamma_1 z)
	\left[ 1 - \epsilon(\omega_0) \right]}
	L(\omega_0)
\end{equation}
where
$r^0 (\omega_0)
= \frac{\hbar \omega ^3}{\pi^2 c^3}$
is the ratio of the Einstein's coefficients in a vacuum.
Fig.~\ref{AB} shows
$r(\omega,z)$
as a function of $\omega$.
From 0 to approximately 35 nm, the ratio decreases, mainly because the (total)
spontaneous emission rate $A_{21}^{(i)}$ decreases. Above 35 nm, the ratio increases,
as $B_{21}$ decreases, because of the exponential decay of the surface plasmon field
away from this interface. Note that for a given frequency $\omega$,
$r(\omega_0, z)$
rises at lower distances for $\omega$ closer to the surface plasmon asymptote frequency.

These results can be used to calculate the amount of power that undergoes
stimulated emission of surface plasmons
in a gain medium in close vicinity of the silver interface.
We consider a parallel beam of surface plasmons.
and suppose that they are excited via a grating or a prism by a He-Ne laser whose emission
has a linewidth of about $\Delta \omega = 10$ MHz centered around $\omega_0 =2$ eV,
and that they
carry $P=1$ mW of power per $\mu$m.
The spectral power at maximum, assuming a lorentzian profile,
is given by $P_{\omega} (\omega_0) = P/ \pi \Delta\omega$, and 
the associated spectral energy per unit surface is
$W(\omega_0) = P_{\omega} (\omega_0) / v_g = \frac{P}{\pi \Delta\omega \, v_g}$ where
$v_g = \frac{\mathrm{d}K}{\mathrm{d}\omega}$ is the group velocity of surface plasmons,
close to $c$ below the asymptote frequency.
The spectral energy per unit surface at maximum of these surface plasmons is then
$W(\omega_0) \approx \frac{P}{\pi \Delta\omega\, c} \approx 10^3\, \hbar .$nm$^2$.
This value is far above those of Fig.~\ref{AB}: stimulated emission
in a freely propagating surface plasmon beam is several
orders of magnitude higher than spontaneous emission.

\section{Conclusion}

In this paper, we have extended previous work on quantization of surface plasmons by introducing a formalism that can use experimental values of the dielectric constant instead of using a specific model for the free electron gas. The key step is the derivation of the energy of a surface plasmon in a dispersive \mbox{non-lossy} medium. The standard quantization scheme in Coulomb's gauge yields the quantum form of the field. This scheme can be extended in a straigthforward way to thin metallic films. To illustrate the formalism, we have derived the spontaneous emission rate of surface plasmons by a two-level system placed close to an interface supporting surface waves as well as  Einstein's coefficients. This quantized theory of surface plasmon will be useful to analyse specific quantum effects such as antibunching, single plasmon interference, quantum coherence properties, but also to derive the interactions of surface waves with other quantum objects, as quantum wells for example.

\appendix
\section{Dielectric constant model} \label{modeleps}
\begin{figure}[htbp]
 \includegraphics[width=8cm]{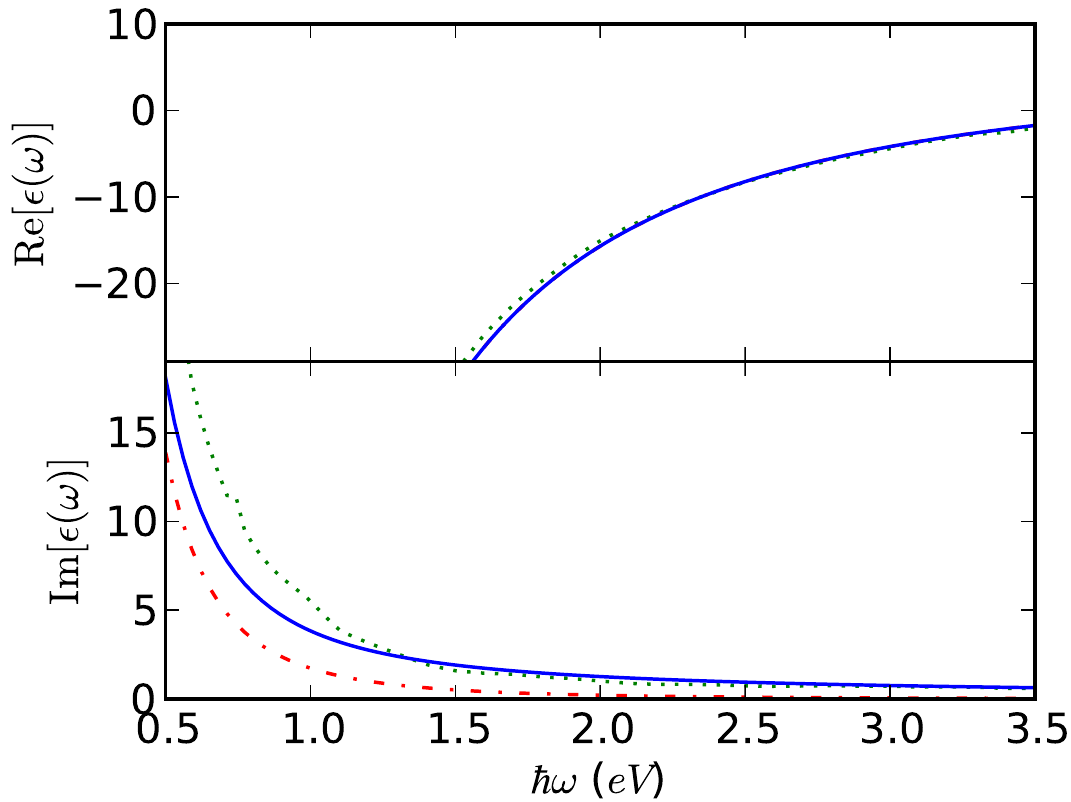}
 \caption{Real and imaginary part of the dielectric constant for silver. The experimental data from Ref.~\onlinecite{Palik} in green dotted line, Drude's model from Ref.~\onlinecite{Oulton} in red and the fit used in this paper in blue. \label{constdielfit}}
 \end{figure}

When cases with losses are considered in this paper, we must sometimes consider the case of complex frequencies.
An analytical model for the dielectric constant is needed
to evaluate $\epsilon(\omega)$ when $\omega$ is complex.
The dielectric constant model which has been used in this paper is the following:
\begin{equation}\label{epseq}
\epsilon(\omega) = \epsilon_{\infty}-\frac{\omega_p^2}{\omega^2+i\beta\omega} + i\frac{\sigma}{\epsilon_0\omega}
\end{equation}
where we take the values $\epsilon_{\infty}=5$, $\hbar\omega_p=9.1$~eV and $\hbar\beta=0.021$~eV of Drude's model given by Ref.~\onlinecite{Oulton}. We add also a conductivity term to have a better fit of the imaginary part of the dielectric constant, we take $\hbar\sigma/\epsilon_0=1.8$~eV.
Fig.~\ref{constdielfit} compares the experimental data\cite{Palik} and both Drude's model from Ref.~\onlinecite{Oulton} and the model used in this paper in Eq.(\ref{epseq}).

\section{Derivation of the energy}\label{energy}

In this section, we focus on the derivation of the electromagnetic energy associated with surface waves. The main idea is to derive the work done by an external operator to build adiabatically the field amplitude in a \mbox{non-lossy} medium.
The energy balance from time $t=0$ to $t=T$ reads:
\begin{equation}
	\label{ene1}
	U = \int _0 ^T \mathrm{d}t \int \mathrm{d}^3 \mathbf{r}
		\left[
			\mathbf{E}\frac{\partial\mathbf{D}}{\partial t}+\mathbf{H}\frac{\partial\mathbf{B}}{\partial t}
		\right]
	= \int _0 ^T \mathrm{d}t \int \mathrm{d}^3\mathbf{r}
		\ (-\mathbf{j}.\mathbf{E})
\end{equation}
in which $\int_0^T\mathrm{d}t\int\mathrm{d}^3\mathbf{r}\ (-\mathbf{j}.\mathbf{E})$ is the operator's work on the system between $t=0$ and $t=T$.
$T$ must be large for this work to be adiabatic. $U$ then does not depend on $T$.
In other words, this is also the total electromagnetic energy of the system. Due to the exponential decrease along the $z$-axis, the contribution of the Poynting vector, which should appear in the left term of equation~(\ref{ene1}), drops to zero. We want hence to derive the first term of equation~(\ref{ene1}) to obtain the electromagnetic energy.

The first step is to introduce time-dependent amplitudes in Eq.~\eqref{expA}: $A_{\mathbf{K}}$ is then replaced by
$A_{\mathbf{K}}(t)$ in which $A_{\mathbf{K}}(t)=A_{\mathbf{K}}\times f(t)$. This accounts for operator's work.
For the sake of convenience, we take a $2T$-periodic function for which $f(0)=f(2T)=0$ and $f(T)=1$, so that we can write $f(t)=\sum_n f_n \exp\left(i 2\pi n \frac{t}{2T}\right)$. $T$ is the typical time of variation of the amplitude $A_{\mathbf{K}}(t)$. $T$ has to be taken sufficiently large to consider the work done by the operator as adiabatic. (For instance $f(t)=\sin\left(\pi \frac{t}{2T}\right)$.) Let us first derive the energy in the material medium denoted medium $2$ for $z<0$. We have:

\begin{widetext}
\begin{equation}
	\label{expE1}
	\mathbf{E}(\mathbf{r},t) =
		-\partial_t \mathbf{A} (\mathbf{r},t)
		=
		-\sum_{\mathbf{K}} \exp(i \mathbf{K}. \mathbf{r}) \mathbf{u} _{\mathbf{K}} (z)
				[-i\omega  A_{\mathbf{K}} (t) + \partial_t  A_{\mathbf{K}} (t)  ] \exp(-i\omega t)
		+ c.c.
\end{equation}
Using $A_{\mathbf{K}} (t) = A_{\mathbf{K}} f(t)$ and the Fourier series expansion of $f(t)$ given above,
the field can be cast
as a sum of terms varying as $\exp[-i(\omega + \frac{\pi n}{T})t]$
\begin{equation}
	\label{expE2}
	\mathbf{E}(\mathbf{r},t) =
		\sum_{\mathbf{K}} \exp(i \mathbf{K}. \mathbf{r}) \mathbf{u} _{\mathbf{K}} (z)
					A_{\mathbf{K}} \sum_n i \left( \omega  - \frac{\pi n}{T} \right) f_n \exp\left[ i \left(\frac{\pi n}{T} - \omega \right) t \right]
		+ c.c.
\end{equation}
Hence the displacement vector
\begin{equation}
	\label{expD}
	\mathbf{D}(\mathbf{r},t) =
		\epsilon_0 \sum_{\mathbf{K}} \exp(i \mathbf{K}. \mathbf{r}) \mathbf{u} _{\mathbf{K}} (z)
					A_{\mathbf{K}} \sum_n i \left( \omega - \frac{\pi n}{T} \right) f_n
					\, \epsilon _j \left(\omega - \frac{\pi n}{T} \right)
					\exp\left[ i \left(\frac{\pi n}{T} - \omega \right) t \right]
		+ c.c.
\end{equation}
and its time derivative
\begin{equation}
	\label{expdtD}
	\partial _t \mathbf{D}(\mathbf{r},t) =
		\epsilon_0 \sum_{\mathbf{K}} \exp(i \mathbf{K}. \mathbf{r}) \mathbf{u} _{\mathbf{K}} (z)
					A_{\mathbf{K}} \sum_n \left( \omega - \frac{\pi n}{T} \right) ^2 f_n
					\, \epsilon _j \left(\omega - \frac{\pi n}{T} \right)
					\exp\left[ i \left(\frac{\pi n}{T} - \omega \right) t \right]
		+ c.c.
\end{equation}
As $1/T \ll \omega$,
we have $(\omega - \frac{\pi n}{T})^2 \epsilon_j (\omega - \frac{\pi n}{T}) \approx \omega ^2 \epsilon_j(\omega) - \frac{\pi n}{T} \frac{\mathrm{d}[ \omega^2 \epsilon_j(\omega)]}{\mathrm{d} \omega}$.
Taking the inverse of the Fourier series expansions in Eq.~\eqref{expdtD}, we get
\begin{equation}
	\partial _t \mathbf{D}(\mathbf{r},t) =
		\epsilon_0 \sum_{\mathbf{K}} \exp(i \mathbf{K}. \mathbf{r}) \mathbf{u} _{\mathbf{K}} (z)
					\left\lbrace
						\omega ^2 \epsilon_j(\omega) A_{\mathbf{K}} (t)
						+ i\frac{\mathrm{d}[ \omega^2 \epsilon_j(\omega)]}{\mathrm{d} \omega} \partial _t A_{\mathbf{K}}(t)
					\right\rbrace
					\exp( -i \omega t )
		+ c.c.
\end{equation}
Hence
\begin{multline}
	\mathbf{E}(\mathbf{r},t) .\partial _t \mathbf{D}(\mathbf{r},t) = \\
		-\epsilon_0
		\sum_{\mathbf{K},\mathbf{K}'}
			\exp[i (\mathbf{K} + \mathbf{K}'). \mathbf{r}]
			\mathbf{u} _{\mathbf{K}} (z)
			.\mathbf{u} _{\mathbf{K}'} (z)
			[-i\omega  A_{\mathbf{K}} (t) + \partial_t  A_{\mathbf{K}} (t) ]
			\left\lbrace
						\omega ^{\prime 2} \epsilon_j(\omega ') A_{\mathbf{K} '} (t)
						+ i\frac{\mathrm{d}[ \omega ^{\prime 2} \epsilon_j(\omega ')]}{\mathrm{d} \omega '} \partial _t A_{\mathbf{K}'}(t)
			\right\rbrace
			\exp[-i(\omega + \omega') t] \\
		-\epsilon_0
		\sum_{\mathbf{K},\mathbf{K}'}
			\exp[i (\mathbf{K} - \mathbf{K}'). \mathbf{r}]
			\mathbf{u} _{\mathbf{K}} (z)
			.\mathbf{u} ^\ast _{\mathbf{K}'} (z)
			[-i\omega  A_{\mathbf{K}} (t) + \partial_t  A_{\mathbf{K}} (t) ]
			\left\lbrace
						\omega ^{\prime 2} \epsilon_j(\omega ') A ^\ast _{\mathbf{K} '} (t)
						- i\frac{\mathrm{d}[ \omega ^{\prime 2} \epsilon_j(\omega ')]}{\mathrm{d} \omega '} \partial _t A ^\ast _{\mathbf{K}'}(t)
			\right\rbrace
			\exp[-i(\omega - \omega') t] \\
		+ c.c.
\end{multline}
Integrating this term over the surface $S$ and using $\int \mathrm{d}x \int \mathrm{d}y \exp(i(\mathbf{K} - \mathbf{K}').\mathbf{r}) = S \delta _{\mathbf{K}, \mathbf{K}'}$ ($\delta$ is the Kroenecker symbol, which verifies
$\delta _{\mathbf{K}, \mathbf{K}'} = 1$ if $\mathbf{K} = \mathbf{K}'$, $\delta _{\mathbf{K}, \mathbf{K}'} = 0$ else), we find
\begin{multline}
	\label{intED}
	\int \! \mathrm{d}x \! \int \! \mathrm{d}y \, \mathbf{E}(\mathbf{r},t) .\partial _t \mathbf{D}(\mathbf{r},t) = \\
		-\epsilon_0 S
		\sum_{\mathbf{K}}
			\mathbf{u} _{\mathbf{K}} (z)
			.\mathbf{u} _{-\mathbf{K}} (z)
			[-i\omega  A_{\mathbf{K}} (t) + \partial_t  A_{\mathbf{K}} (t) ]
			\left\lbrace
						\omega ^2 \epsilon_j(\omega) A_{-\mathbf{K}} (t)
						+ i\frac{\mathrm{d}[ \omega ^2 \epsilon_j(\omega)]}{\mathrm{d} \omega} \partial _t A_{-\mathbf{K}}(t)
			\right\rbrace
			\exp(-2i \omega t) \\
		-\epsilon_0 S
		\sum_{\mathbf{K}}
			\mathbf{u} _{\mathbf{K}} (z)
			.\mathbf{u} ^\ast _{\mathbf{K}} (z)
			[-i\omega  A_{\mathbf{K}} (t) + \partial_t  A_{\mathbf{K}} (t) ]
			\left\lbrace
						\omega ^2 \epsilon_j(\omega) A ^\ast _{\mathbf{K}} (t)
						- i\frac{\mathrm{d}[ \omega ^2 \epsilon_j(\omega)]}{\mathrm{d} \omega} \partial_t A ^\ast _{\mathbf{K}}(t)
			\right\rbrace \\
		+ c.c.
\end{multline}
The first term on the right hand side of Eq.~\eqref{intED} can be expanded and integrated from
$t=0$ to $t=T$. For $T$ sufficiently large, all the terms vanish except the one proportional
to $A_{\mathbf{K}}(t) A_{-\mathbf{K}}(t)$ which gives the first term on the right hand side of
Eq.~\eqref{intT_ED}. The second term on the right hand side of Eq.~\eqref{intED} can be expanded
and integrated from $t=0$ to $t=T$ too. The terms proportional to
$A_{\mathbf{K}} A ^* _{\mathbf{K}}$ and $\partial_t A_{\mathbf{K}} \partial_t A ^* _{\mathbf{K}}$
are pure imaginary and give no contribution once added to their complex conjugate and opposite.
The terms proportional to 
$A_{\mathbf{K}} \partial_t A ^* _{\mathbf{K}}$ and $\partial_t A_{\mathbf{K}} \  A ^* _{\mathbf{K}}$,
once added to their complex conjugate, both vary as $\partial _t \lvert A_{\mathbf{K}} \rvert ^2 $
and are easily integrated to give the last term on the right hand side of Eq.~\eqref{intT_ED}.
\begin{multline}
	\label{intT_ED}
	\int _0 ^T \! \mathrm{d}t
	\int \! \mathrm{d}x \! \int \! \mathrm{d}y \, \mathbf{E}(\mathbf{r},t) .\partial _t \mathbf{D}(\mathbf{r},t) = \\
		\epsilon_0 S
		\sum_{\mathbf{K}}
			\mathbf{u} _{\mathbf{K}} (z)
			.\mathbf{u} _{-\mathbf{K}} (z)
			i \omega ^3 \epsilon_j(\omega)
			\int _0 ^T \! \mathrm{d}t \left[ A_{\mathbf{K}} (t) A_{-\mathbf{K}} (t) \exp(-2i \omega t) \right]
		+ c.c.\\
		+\epsilon_0 S
		\sum_{\mathbf{K}}
			\lvert \mathbf{u} _{\mathbf{K}} (z) \rvert ^2
			\omega ^2
			\frac{\mathrm{d}[ \omega \epsilon_j(\omega)]}{\mathrm{d} \omega}
				 \lvert A_{\mathbf{K}} \rvert ^2 .
\end{multline}
Let us now calculate the energy associated with the term $\mathbf{H}.\frac{\partial \mathbf{B}}{\partial t} = \frac{1}{2\mu_0} \frac{\partial \mathbf{B}^2}{\partial t}$
in Eq.~\eqref{ene1}.
\begin{eqnarray}
	\mathbf{H} (\mathbf{r}, t) &=& \frac{1}{\mu_0} \nabla \times \mathbf{A} (\mathbf{r}, t)
		= i \sum_{\mathbf{K}} \exp (i\mathbf{K}.\mathbf{r}) \mathbf{b}_{\mathbf{K}} (z)
				A_{\mathbf{K}} (t) \exp(-i\omega t) + c.c. \\
	\partial_t \mathbf{B} (\mathbf{r}, t) &=& \partial _t \nabla \times \mathbf{A} (\mathbf{r}, t)
		= i \sum_{\mathbf{K}} \exp (i\mathbf{K}.\mathbf{r}) \mathbf{b}_{\mathbf{K}} (z)
				\left[ \partial _t A_{\mathbf{K}} (t) - i \omega  A_{\mathbf{K}} (t) \right] \exp(-i\omega t) + c.c.
\end{eqnarray}
Hence
\begin{multline}
	\mathbf{H} (\mathbf{r}, t) . \partial_t \mathbf{B} (\mathbf{r}, t)
		=
		  - \frac{1}{\mu_0} \sum_{\mathbf{K},\mathbf{K}'}
			\exp [i(\mathbf{K}+\mathbf{K}').\mathbf{r}]
			\mathbf{b}_{\mathbf{K}} (z)
				. \mathbf{b}_{\mathbf{K}'} (z)
			A_{\mathbf{K}} (t)
				\left[ \partial _t A_{\mathbf{K}'} (t) - i \omega '  A_{\mathbf{K}'} (t) \right]
			\exp[-i(\omega+\omega') t] \\
		  + \frac{1}{\mu_0} \sum_{\mathbf{K},\mathbf{K}'}
			\exp [i(\mathbf{K}-\mathbf{K}').\mathbf{r}]
			\mathbf{b}_{\mathbf{K}} (z)
				. \mathbf{b} ^* _{\mathbf{K}'} (z)
			A_{\mathbf{K}} (t)
				\left[ \partial _t A ^* _{\mathbf{K}} (t) + i \omega  A ^* _{\mathbf{K}} (t) \right]
			\exp[-i(\omega-\omega') t]
		+ c.c.
\end{multline}
Integration over the surface $S$ as previously for $\mathbf{E}. \partial_t \mathbf{D}$ yields:
\begin{multline}
	\label{HdtB}
	\int \! \mathrm{d}x \int \! \mathrm{d}y \ 
			\mathbf{H} (\mathbf{r}, t) . \partial_t \mathbf{B} (\mathbf{r}, t)
		=
		  - \frac{S}{\mu_0} \sum_{\mathbf{K}}
			\mathbf{b}_{\mathbf{K}} (z)
				. \mathbf{b}_{-\mathbf{K}} (z)
			A_{\mathbf{K}} (t)
				\left[ \partial _t A_{-\mathbf{K}} (t) - i \omega  A_{-\mathbf{K}} (t) \right]
			\exp(-2i\omega t) \\
		  + \frac{S}{\mu_0}  \sum_{\mathbf{K}}
			\lvert \mathbf{b}_{\mathbf{K}} (z) \rvert ^2
			A_{\mathbf{K}} (t)
				\left[ \partial _t A ^* _{\mathbf{K}} (t) + i \omega  A ^* _{\mathbf{K}} (t) \right]
		+ c.c.
\end{multline}
The first term on the right hand side of Eq.~\eqref{HdtB} can be expanded and integrated
from $t=0$ to $t=T$. For $T$ sufficiently large, the term proportional to
$A_{\mathbf{K}} (t) \partial _t A_{-\mathbf{K}} (t)$ vanishes, and we get the first
term on the right hand side of Eq.~\eqref{intT_HdtB}.
The second term on the right hand side of Eq.~\eqref{HdtB} can be expanded and integrated
from $t=0$ to $t=T$ too. The term proportional to 
$A_{\mathbf{K}} (t) A ^* _{\mathbf{K}} (t)$ is a pure imaginary, and gives no contribution
once added to its complex conjugate. The term varying as $A_{\mathbf{K}} (t) \partial _t A ^* _{\mathbf{K}} (t)$, once added to its complex conjugate, varies as $\partial _t \lvert A _{\mathbf{K}} (t) \rvert ^2$ and is easily integrated to give the last term on the right hand side of Eq.~\eqref{intT_HdtB}.
\begin{multline}
	\label{intT_HdtB}
	\int _0 ^T \mathrm{d}t \  \int \! \mathrm{d}x \int \! \mathrm{d}y \ 
			\mathbf{H} (\mathbf{r}, t) . \partial_t \mathbf{B} (\mathbf{r}, t)
		= \\
		   \frac{S}{\mu_0}  \sum_{\mathbf{K}}
			i \mathbf{b}_{\mathbf{K}} (z)
				. \mathbf{b}_{-\mathbf{K}} (z)
			\int _0 ^T \mathrm{d}t \  \left[ A_{\mathbf{K}} (t)
				\omega  A_{-\mathbf{K}} (t)
			\exp(-2i\omega t) \right]
		+ c.c. \\
		  + \frac{S}{\mu_0}  \sum_{\mathbf{K}}
			\lvert \mathbf{b}_{\mathbf{K}} (z) \rvert ^2
			\lvert A_{\mathbf{K}} \rvert ^2 .
\end{multline}
We now add Eqs.~\eqref{intT_ED} and \eqref{intT_HdtB}. The first terms on the right hand sides
of these two equations cancel each other (so do their complex conjugates), so that we get
\begin{multline}
	\label{intTxy}
	\int _0 ^T \mathrm{d}t \  \int \! \mathrm{d}x \int \! \mathrm{d}y \ 
			\left[
				\mathbf{E}(\mathbf{r},t) .\partial _t \mathbf{D}(\mathbf{r},t) + 
				\mathbf{H} (\mathbf{r}, t) . \partial_t \mathbf{B} (\mathbf{r}, t)
			\right]
		=
		   \sum_{\mathbf{K}} \epsilon_0 S
			\omega ^2
			\left[
			\frac{\mathrm{d}[ \omega \epsilon_j(\omega)]}{\mathrm{d} \omega}
			\lvert \mathbf{u} _{\mathbf{K}} (z) \rvert ^2
			+
			\frac{c^2}{\omega ^2}
			\lvert \mathbf{b}_{\mathbf{K}} (z) \rvert ^2
			\right]
			\lvert A_{\mathbf{K}} \rvert ^2 .
\end{multline}
Before integrating this expression over $z$, let us remark that it can be cast
under a somewhat clearer form
\begin{multline}
	\label{intTxy_simple}
	\int _0 ^T \mathrm{d}t \  \int \! \mathrm{d}x \int \! \mathrm{d}y \ 
			\left[
				\mathbf{E}(\mathbf{r},t) .\partial _t \mathbf{D}(\mathbf{r},t) + 
				\mathbf{H} (\mathbf{r}, t) . \partial_t \mathbf{B} (\mathbf{r}, t)
			\right]
		=
		   S \sum_{\mathbf{K}} 
			\left[
			\epsilon_0 \frac{\mathrm{d}[ \omega \epsilon_j(\omega)]}{\mathrm{d} \omega}
			\lvert \mathbf{E} _{\mathbf{K}} (z) \rvert ^2
			+
			\frac{1}{\mu_0}
			\lvert \mathbf{B}_{\mathbf{K}} (z) \rvert ^2
			\right].
\end{multline}
where $\mathbf{E} _{\mathbf{K}} (z) = i\omega A_{\mathbf{K}} \mathbf{u}_{\mathbf{K}} (z)$
and
$\mathbf{B}_{\mathbf{K}} (z) = i A_{\mathbf{K}} \mathbf{b}_{\mathbf{K}} (z)$
so that $\mathbf{E} (\mathbf{r}, t) = \sum_{\mathbf{K}} \exp(i\mathbf{K}.\mathbf{r}) \mathbf{E} _{\mathbf{K}} (z) \exp(-i\omega t) + c.c.$
and
$\mathbf{B} (\mathbf{r}, t) = \sum_{\mathbf{K}} \exp(i\mathbf{K}.\mathbf{r}) \mathbf{B} _{\mathbf{K}} (z) \exp(-i\omega t) + c.c.$. Eq.~\eqref{intTxy_simple} gives the energy per unit length along the $z$ direction of
surface plasmons. $\epsilon_0 S \frac{\mathrm{d}[ \omega \epsilon_j(\omega)]}{\mathrm{d} \omega}
			\lvert \mathbf{E} _{\mathbf{K}} (z) \rvert ^2$ and
$
			\frac{1}{\mu_0} S
			\lvert \mathbf{B}_{\mathbf{K}} (z) \rvert ^2
$
are then the electric and magnetic contribution of each mode $\mathbf{K}$ to the former
energy per unit length.

We now would like to integrate Eq.~\eqref{intTxy} over $z$. In both half spaces $z>0$ and $z<0$
($j=1$, $2$), we have
$\lvert \mathbf{u}_{\mathbf{K}} (z) \rvert ^2 = \frac{1}{L(\omega)} \exp[- 2 \mathrm{Im}(\gamma_j) z] \left( 1 + \frac{K^2}{\lvert \gamma_j \rvert ^2} \right) $,
$\mathbf{b}_{\mathbf{K}} (z) = (K\mathbf{\hat{K}} + \gamma_j \mathbf{\hat{z}}) \times \frac{1}{\sqrt{L(\omega)}} \exp(i\gamma_j z) (\mathbf{\hat{K}} - \frac{K}{\gamma_j} \mathbf{\hat{z}})
= \frac{1}{\sqrt{L(\omega)}} \exp(i\gamma_j z) \frac{\epsilon_j}{\gamma_j} \frac{\omega ^2}{c^2}  \mathbf{\hat{z}} \times \mathbf{\hat{K}} $ (we used the property $\mathbf{\hat{z}}.\mathbf{\hat{K}} = 0$), hence
$ \lvert \mathbf{b}_{\mathbf{K}} (z) \rvert ^2 = \frac{1}{L(\omega)} \exp[-2 \mathrm{Im}(\gamma_j) z] \left\lvert \frac{\epsilon_j}{\gamma_j} \right\rvert ^2 \frac{\omega ^4}{c^4}$.
Now writing
\begin{subequations}
\label{intZ}
\begin{eqnarray}
	\int _{-\infty} ^\infty \! \mathrm{d}z \  
		\lvert \mathbf{u} _{\mathbf{K}} (z) \rvert ^2
			\frac{\mathrm{d}[ \omega \epsilon_j(\omega)]}{\mathrm{d} \omega}
	&=&
		\frac{1}{L(\omega)} \sum _{j=1,2}
			\frac{1}{2 \lvert \gamma_j \rvert} \left( 1 + \frac{K^2}{\lvert \gamma_j \rvert ^2} \right)
			\frac{\mathrm{d}[ \omega \epsilon_j(\omega)]}{\mathrm{d} \omega}
		\\
	\int _{-\infty} ^\infty \! \mathrm{d}z \  
		\lvert \mathbf{b}_{\mathbf{K}} (z) \rvert ^2
	&=&
		\frac{1}{L(\omega)} \sum_{j=1,2}
			\frac{1}{2 \lvert \gamma_j \rvert} \left\lvert \frac{\epsilon_j}{\gamma_j} \right\rvert ^2 \frac{\omega ^4}{c^4}
\end{eqnarray}
\end{subequations}
Integrating Eq.~\eqref{intTxy} over $z$ using Eqs.~\eqref{intZ}, we get
\begin{multline}
	\label{U_L}
	U = \int _0 ^T \mathrm{d}t \  \int \! \mathrm{d}^3 \mathbf{r} \ 
			\left[
				\mathbf{E}(\mathbf{r},t) .\partial _t \mathbf{D}(\mathbf{r},t) + 
				\mathbf{H} (\mathbf{r}, t) . \partial_t \mathbf{B} (\mathbf{r}, t)
			\right]
		= \\
		   \sum_{\mathbf{K}} \epsilon_0 S
			\omega ^2
			\frac{1}{L(\omega)} \sum_{j=1,2}
				\frac{1}{2 \lvert \gamma_j \rvert}
			\left[
				\left( 1 + \frac{K^2}{\lvert \gamma_j \rvert ^2} \right)
				\frac{\mathrm{d}[ \omega \epsilon_j(\omega)]}{\mathrm{d} \omega}
			+
				\left\lvert \frac{\epsilon_j}{\gamma_j} \right\rvert ^2 \frac{\omega ^2}{c^2}
			\right]
			\lvert A_{\mathbf{K}} \rvert ^2 .
\end{multline}
We now use the degree of freedom to set $L(\omega)$ as we wish, to simplify this equation.
We set
\begin{equation}
		\label{NormalizationL0}
			L(\omega) = \frac{1}{2} \sum_{j=1,2}
				\frac{1}{2 \lvert \gamma_j \rvert}
			\left[
				\left( 1 + \frac{K^2}{\lvert \gamma_j \rvert ^2} \right)
				\frac{\mathrm{d}[ \omega \epsilon_j(\omega)]}{\mathrm{d} \omega}
			+
				\left\lvert \frac{\epsilon_j}{\gamma_j} \right\rvert ^2 \frac{\omega ^2}{c^2}
			\right].
\end{equation}
Using Eq.~\eqref{reldisp}, the definition of $\gamma_j$, and $\epsilon(\omega) \leq -1$ at
the frequencies of surface plasmon of a single interface,
Eq.~\eqref{NormalizationL0}
writes
\begin{equation}
		\label{NormalizationL}
			L(\omega) =
				\frac{-\epsilon(\omega)}{2 \lvert \gamma_1 \rvert}
			+
				\frac{1}{4 \lvert \gamma_2 \rvert}
			\left[
				\frac{1- \epsilon(\omega)}{- \epsilon(\omega) }
				\frac{\mathrm{d}[ \omega \epsilon(\omega)]}{\mathrm{d} \omega}
			-
				 1-\epsilon(\omega)
			\right]
\end{equation}
This is equivalent to set
\begin{equation}
		\label{NormalizationU0}
		\int _{-\infty} ^\infty \! \mathrm{d}z \ 
		 \frac{1}{2} \left[
			\frac{\mathrm{d}[ \omega \epsilon_j(\omega)]}{\mathrm{d} \omega}
			\lvert \mathbf{u} _{\mathbf{K}} (z) \rvert ^2
			+
			\frac{c^2}{\omega ^2}
			\lvert \mathbf{b}_{\mathbf{K}} (z) \rvert ^2
		\right] = 1
\end{equation}
(the term inside the brackets comes from Eq.~\eqref{intTxy}).
Using the expressions of $\lvert \mathbf{u} _{\mathbf{K}} (z) \rvert ^2$
and $\lvert \mathbf{b} _{\mathbf{K}} (z) \rvert ^2$ given above,
Eq.~\eqref{reldisp} and the definition of $\gamma_j$,
this gives a normalization condition on $\mathbf{u}_{\mathbf{K}} (z)$
\begin{equation}
		\label{NormalizationU}
		\int _{-\infty} ^\infty \! \mathrm{d}z \ 
		 \frac{1}{2} \left[
			\frac{\mathrm{d}[ \omega \epsilon_j(\omega)]}{\mathrm{d} \omega}
			+
			\lvert \epsilon _j (\omega) \rvert
			\frac{\lvert 1 + \epsilon (\omega) \rvert}{1 + \lvert \epsilon (\omega) \rvert}
		\right]
			\lvert \mathbf{u} _{\mathbf{K}} (z) \rvert ^2
			= 1.
\end{equation}
Eqs.~\eqref{NormalizationU0} and \eqref{NormalizationU} can also be written
\begin{equation}
		\label{NormalizationUc}
		\int _{-\infty} ^\infty \! \mathrm{d}z \ 
		 \left[
			\frac{\epsilon_0}{2} 
			\frac{\mathrm{d}[ \omega \epsilon_j(\omega)]}{\mathrm{d} \omega}
			\lvert \mathbf{E} _{\mathbf{K}} (z) \rvert ^2
			+
			\frac{1}{2\mu_0} 
			\lvert \mathbf{B}_{\mathbf{K}} (z) \rvert ^2
		\right] = \epsilon_0 \omega ^2 \lvert A_{\mathbf{K}} \rvert^2 ,
\end{equation}
with the notations used in Eq.~\eqref{intTxy_simple}.

With this choice of $L(\omega)$ or this normalization condition of $\mathbf{u} _{\mathbf{K}} (z)$,
Eq.~\eqref{U_L} can be written:
\begin{equation}
	\label{U_f}
	U
		=
		   \sum_{\mathbf{K}} \epsilon_0 S
			\omega ^2
			2 \lvert A_{\mathbf{K}} \rvert ^2 .
\end{equation}

\end{widetext}

\section{Quantum calculation of the surface plasmon emission rate of a dipole}
\label{quantumcalc}
We follow the same steps as in the derivation of the photon emission rate by a two-level system in a
vacuum.
The fundamental and excited states of the two-level system are denoted $\ket{1}$ and $\ket{2}$ respectively, associated with energies $E_1$ and $E_2$. We define a circular frequency $\omega_0$, so that $E_2-E_1=\hbar\omega_0$.
The two-level system is initially in its excited state, and there are $n_\mathbf{K}$ surface plasmons so that the global  initial state can be written: $\ket{i}=\ket{2,n_\mathbf{K}}$.
In the final state, the dipole is in its fundamental state and a surface plasmon has been created  in a mode $\mathbf{K}$.
We denote the final global state $\ket{f}=\ket{1,n_\mathbf{K}+1}$.
The interaction Hamiltonian is $-\widehat{\mathbf{D}} .\widehat{\mathbf{E}}$, where $\widehat{\mathbf{D}}$ is the electric-dipole moment operator and
$\widehat{\mathbf{E}}$ the quantum electric field operator introduced in section \ref{sec_quantization}, at the position of the emitter.
The emission rate (inverse of the lifetime $\tau$ of the excited state) is given by Fermi's golden rule:
\begin{equation}\label{Fermi}
 \gamma=
\frac{2\pi}{\hbar}
	\sum_{f} |\bra{f}\hat{\mathbf{D}}.\hat{\mathbf{E}} \ket{i}|^2\delta\left(E_2-E_1 - \hbar\omega\right).
\end{equation}
This expression can be rewritten as a sum over the modes $\mathbf{K}$:
\begin{equation}
\label{Tau}
 \gamma=
	\frac{2\pi}{\hbar}
	\sum_{\mathbf{K}} M_{\mathbf{K}}
		\delta\left(E_2-E_1 - \hbar\omega\right),
\end{equation}
where
$M_{\mathbf{K}} = |\bra{1,n_\mathbf{K}+1}\hat{\mathbf{D}}.\hat{\mathbf{E}} \ket{2, n_{\mathbf{K}}}|^2$.
We note $\bra{2}\widehat{\mathbf{D}}\ket{1}= \mathbf{D}_{12} $.

Using the former
expression and 
Eqs.~(\ref{a+}), (\ref{a}) and (\ref{Equantif}),
we obtain the following matrix element:
\begin{equation}\label{EltMat}
\mathrm{M}_{\mathbf{K}} =
\frac{\hbar\omega}{2\epsilon_0 S}
\left\lvert \mathbf{D}_{12} . \mathbf{u} _{1, \mathbf{K}} (z) \right\rvert ^2
				\ (n_{\mathbf{K}}+1).
\end{equation}
In this equation, the $n_\mathbf{K}$ term stands for the stimulated emission and the constant term 1 accounts for the spontaneous emission. This section is devoted to 
the spontaneous emission so that
we do not consider the term associated to
$n_{\mathbf{K}}$.

Using Eqs.~\eqref{Tau} and \eqref{EltMat}, substituting a continuous sum over the vectors
$\mathbf{K}$ in polar coordinates to the discrete sum $\sum_{\mathbf{K}}$,
and now writing $\gamma_{spont}$ instead of $\gamma$
we get
\begin{multline}
\gamma_{spont} =
	\frac{2\pi}{\hbar}
	\int _{0} ^{\infty} \mathrm{d} K \, K
		\frac{S}{(2\pi)^2} \ 
		\frac{\hbar\omega}{2\epsilon_0 S}
		 \delta\left(E_2-E_1 - \hbar\omega\right)
		\\
	\times \int _{0} ^{2\pi} \mathrm{d}\theta \,
		\left\lvert \mathbf{D}_{12}
			. \mathbf{u} _{1, \mathbf{K}} (z) \right\rvert ^2 .
\end{multline}
The integration over the directions $\theta$ of $\mathbf{K}$ is performed using
Eq.\eqref{Uj}:
\begin{multline}
	\int _{0} ^{2\pi} \mathrm{d}\theta \,
		\left\lvert \mathbf{D}_{12}
			. \mathbf{u} _{1, \mathbf{K}} (z) \right\rvert ^2
	= \frac{2 \pi}{L_{eff} (z, \mathrm{d}_{12}, \omega_0)}
\end{multline}
$L_{eff} (z, \mathrm{d}_{12}, \omega_0)$ is defined by Eq.~\eqref{Leff}.
The spontaneous emission rate of surface plasmon can then be cast in the form given by Eq.~\eqref{Rate}.

Averaging $\gamma_{spont}$
over all the possible directions ($\theta$, $\phi$) and all the precession angles $\psi$ of $\mathbf{D}_{12}$,
we get the "total" spontaneous emission rate of surface plasmons
\begin{multline}
	\gamma_{spont,total} (\lvert \mathbf{D}_{12} \rvert, \omega_0, z) = \\
		\frac{1}{8\pi^2}
		\int _0 ^\pi \mathrm{d}\theta \, \sin\theta
		\int _0 ^{2\pi} \mathrm{d}\phi
		\int _0 ^{2\pi} \mathrm{d}\psi \ 
		\gamma_{spont} (\mathbf{R} _{\theta, \phi, \psi} (\mathbf{D}_{12}), \omega_0, z)
\end{multline}
where $\mathbf{R} _{\theta, \phi, \psi} (\mathbf{D}_{12})$ is $\mathbf{D}_{12}$
rotated by Euler's angles for nutation, precession and intrinsic rotation $\theta$, $\phi$ and $\psi$ respectively.
The result of the integration over $\theta$, $\phi$ and $\psi$ is given by Eq.~\eqref{RateTotal}.

\section{Derivation of the emission rate of a dipole in the classical lossy case}
\label{rategreen}
The aim of this section is to derive an explicit form of the Purcell factor due to the presence
of surface plasmons by using the Green's tensor approach.
Using Ref.~\onlinecite{PRBAlex}, one can write the surface plasmon contribution to the Green's
tensor
evaluated at the position of the source $\mathbf{r}$:
\begin{multline}
	\label{Gspstart}
	\tensor{\mathbf{G}}_{sp}(\mathbf{r}, \mathbf{r}, \omega) = 
		\int  \frac{\mathrm{d} ^2 \mathbf{K}}{(2\pi)^2}
			\left[
			  \frac{\tensor{\mathbf{f}} (\mathbf{K}, z, z)}{\omega - \omega_{sp}}
			  - \frac{\tensor{\mathbf{f}} ^\ast (-\mathbf{K}, z, z)}{\omega + \omega ^\ast _{sp}}
			\right]
\end{multline}
where $\tensor{\mathbf{f}} (\mathbf{K}, z, z)$ is given by
\begin{equation}
	\label{ftensor}
	\tensor{\mathbf{f}} (\mathbf{K}, z, z) =
		-g(K, \omega_{sp}) (\mathbf{\hat{K}} - \frac{K}{\gamma_1} \mathbf{\hat{z}})
		(\mathbf{\hat{K}} - \frac{K}{\gamma_1} \mathbf{\hat{z}})
		\exp(2i\gamma_1 z)
\end{equation}
as $z>0$,
with
$g(K, \omega_{sp}) = \frac{c^2 \gamma_1 ^2 \epsilon(\omega_{sp})}{\omega _{sp} ^2} R(K, \omega_{sp})$
and
$R ^{-1} (K, \omega_{sp}) = -i  \frac{\partial }{\partial \omega} \left[ \gamma_1 (K, \omega) \epsilon (\omega) - \gamma_2 (K, \omega) \right] \lvert_{\omega = \omega_{sp}}$.
Injecting Eq.~\eqref{Gspstart}
in Eq.~\eqref{Novot}, we obtain:
\begin{widetext}
\begin{equation}
	F_{P,cl} (\mathbf{d}_{12}, \omega, z) =
		\frac{6 \pi c}{\omega} \ 
		\int  \frac{\mathrm{d} ^2 \mathbf{K}}{(2\pi)^2}
		\ \mathrm{Im} \Bigg[ 
			  \frac{\mathbf{d} ^\ast _{12} . \tensor{\mathbf{f}} (\mathbf{K}, z, z) \mathbf{d}_{12}}{\omega - \omega_{sp}}
			  - \frac{\mathbf{d} ^\ast _{12} . \tensor{\mathbf{f}} ^\ast (-\mathbf{K}, z, z) \mathbf{d}_{12}}{\omega + \omega ^\ast _{sp}}
		 \Bigg].
\end{equation}
Writing $F_{\mathbf{K}}  (\mathbf{d}_{12}, \omega, z) = \mathbf{d} ^\ast _{12} . \tensor{\mathbf{f}} (\mathbf{K}, z, z) \mathbf{d}_{12} = F_{\mathbf{K}} ' (\mathbf{d}_{12}, \omega, z) + iF_{\mathbf{K}}'' (\mathbf{d}_{12}, \omega, z)$
and
$\rho ' _{\omega_{sp}} (\omega) = \frac{\omega - \omega ' _{sp}}{(\omega - \omega ' _{sp})^2 + \omega_{sp}^{\prime \prime 2}}$, 
$\rho '' _{\omega_{sp}} (\omega) = \frac{\omega '' _{sp}}{(\omega - \omega ' _{sp})^2 + \omega_{sp}^{\prime \prime 2}}$, we find:
\begin{multline}
	\label{RateW_sum}
	F_{P,cl} (\mathbf{d}_{12}, \omega, z) =
		\frac{6 \pi c}{\omega} \ 
		\int  \frac{\mathrm{d} ^2 \mathbf{K}}{(2\pi)^2}
			\bigg[ F_{\mathbf{K}} ' (\mathbf{d}_{12}, \omega, z) \rho '' _{\omega_{sp}} (\omega)
			+ F_{\mathbf{K}}'' (\mathbf{d}_{12}, \omega, z) \rho ' _{\omega_{sp}} (\omega)
			\\
			-F_{\mathbf{K}} ' (\mathbf{d}_{12}, \omega, z) \rho '' _{-\omega ^\ast _{sp}} (\omega)
			+F_{\mathbf{K}}'' (\mathbf{d}_{12}, \omega, z) \rho ' _{-\omega ^\ast _{sp}} (\omega)
		\bigg].
\end{multline}
This expression is to be compared to Eq.~\eqref{Tau} in the non lossy case. Both expressions
are written as sums over the modes $\mathbf{K}$, of the contribution of each mode
to the spontaneous emission rate.
When losses are low, the last two terms on the right hand side of Eq.~\eqref{RateW_sum}
are antiresonant, as $\rho '' _{-\omega ^\ast _{sp}} (\omega)$
and $\rho ' _{-\omega ^\ast _{sp}} (\omega)$ are centered around $-\omega ' _{sp}$. The
second term has also a small contribution as the average value of $\rho ' _{\omega_{sp}} (\omega)$
is $0$. The main contribution comes from the first term, as
$\rho '' _{\omega_{sp}} (\omega)$ goes to $-i\pi \delta (\omega - \omega_{sp})$
in the non lossy limit.
Its extremum value
is
$\frac{-2Q}{\omega '_{sp}}$.
Starting again from Eq.~\eqref{Gspstart},
we write 
\begin{equation}
	\label{Gsp}
	\tensor{\mathbf{G}}_{sp}(\mathbf{r}, \mathbf{r}, \omega) = 
		\frac{1}{(2\pi)^2}
		\int _0 ^\infty \mathrm{d}K \, K
		 \ 
			\left\lbrace
			  \frac{1}{\omega - \omega_{sp}} \int _0 ^{2\pi} \mathrm{d}\theta \ \left[\tensor{\mathbf{f}} (\mathbf{K}, z, z)\right]
			  - \frac{1}{\omega + \omega ^\ast _{sp}} \int _0 ^{2\pi} \mathrm{d}\theta \ \left[\tensor{\mathbf{f}} ^\ast (-\mathbf{K}, z, z)\right]
			\right\rbrace,
\end{equation}
aiming at performing the integral over the directions $\theta$ of $\mathbf{K}$.
We first calculate:
\begin{equation}
	\label{intTheta}
	\int _0 ^{2\pi} \mathrm{d}\theta \ 
		  \left[\tensor{\mathbf{f}} (\mathbf{K}, z, z)\right] = 
	-2\pi g(K, \omega_{sp})
		\tensor{\mathbf{U}}(z, \omega_{sp}),
\end{equation}
where $\tensor{\mathbf{U}}(z, \omega_{sp}) = \exp(2i\gamma_1 z) \left[ \frac{1}{2} (\mathbf{\hat{x}} \mathbf{\hat{x}} + \mathbf{\hat{y}} \mathbf{\hat{y}})
		- \epsilon(\omega _{sp}) \mathbf{\hat{z}} \mathbf{\hat{z}} \right]$.
Inserting Eq.~\eqref{intTheta} in Eq.~\eqref{Gsp} yields:
\begin{equation}
	\label{GspF}
	\tensor{\mathbf{G}}_{sp}(\mathbf{r}, \mathbf{r}, \omega) = 
		\frac{1}{2\pi}
		\int _0 ^\infty \mathrm{d}K \, K
		\ 
			\Bigg[
	\frac{-g(K, \omega_{sp})}{\omega - \omega_{sp}}
		\tensor{\mathbf{U}}(z, \omega_{sp})
	+\frac{g ^\ast (K, \omega_{sp})}{\omega + \omega ^\ast _{sp}}
		\tensor{\mathbf{U}} ^\ast (z, \omega_{sp})
			\Bigg].
\end{equation}
Inserting Eq.~\eqref{GspF} in Eq.~\eqref{Novot},
one finds:
\begin{equation}
	\label{RateW}
	F_{P,cl} (\mathbf{d}_{12}, \omega_0, z) =
		\frac{3 c}{\omega_0} 
		\int _0 ^\infty \mathrm{d}K \, K
		\  \mathrm{Im}
			\Bigg[
			  \frac{-g(K, \omega_{sp})
	}{\omega_0 - \omega_{sp}}
		\mathbf{d} ^\ast _{12} . \tensor{\mathbf{U}}(z,\omega_{sp}) \mathbf{d}_{12}
			  + \frac{g ^\ast (K, \omega_{sp})
	}{\omega_0 + \omega ^\ast _{sp}}
		\mathbf{d} _{12} . \tensor{\mathbf{U}}^\ast (z,\omega_{sp}) \mathbf{d}^\ast _{12}
			\Bigg].
\end{equation}
with
$\mathbf{d} ^\ast _{12} . \tensor{\mathbf{U}}(z,\omega_{sp}) \mathbf{d}_{12}
	= \exp(2i\gamma_1 z)
	  \left[
		\frac{1}{2} \lvert \mathbf{d}_{12,/\!/} \rvert ^2
		- \epsilon(\omega _{sp}) \lvert d_{12,z} \rvert ^2
	  \right]$.
In the non lossy case, $g(K, \omega_{sp})$ and
$\mathbf{d} ^\ast _{12} . \tensor{\mathbf{U}}(z,\omega_{sp}) \mathbf{d}_{12}$
are real, and $\mathrm{Im} \frac{1}{\omega - \omega _{sp}}$ goes to\cite{Novotny}
$-i\pi \delta (\omega - \omega_{sp})$. Hence, assuming $\omega > 0$ and using the
expressions of $\mathbf{d} ^\ast _{12} . \tensor{\mathbf{U}}(z,\omega_{sp}) \mathbf{d}_{12}$ and $g(K, \omega_{sp})$ given above, one gets Eq.~\eqref{Gcl_ll}.
We need the relation
\begin{equation}
	\label{RL}
	R(K, \omega_{sp}) =
	\frac{\omega_{sp}}{K^2}
	\frac{1}{2 L(\omega_{sp})},
\end{equation}
in order to prove that 
$F_P (\mathbf{d}_{12}, \omega_0, z) = F_{P,cl} (\mathbf{d}_{12}, \omega_0, z)$.
Using the definitions of $R(K, \omega_{sp})$ and of $\gamma_j$ given above, we get
\begin{eqnarray}
	\label{R1}
	R ^{-1} (K, \omega_{sp})
		&=& -i \frac{\epsilon\omega/c^2}{\gamma_1}
			-i \gamma_1 \frac{\mathrm{d} \epsilon}{\mathrm{d} \omega }
			+i \frac{\frac{\mathrm{d} \epsilon}{\mathrm{d} \omega }\frac{\omega ^2}{c^2} + 2\omega \epsilon /c^2}{2\gamma_2}
\end{eqnarray}
Inserting $\frac{\mathrm{d} \epsilon}{\mathrm{d} \omega } = \frac{1}{\omega}\left[ \frac{\mathrm{d} \omega \epsilon}{\mathrm{d} \omega } - \epsilon \right]$ in Eq.~\eqref{R1}, and using $\gamma_1 = i\lvert \gamma_1 \rvert$, $\gamma_2 = -i\lvert\gamma_2\rvert$, we find
\begin{eqnarray}
	R ^{-1} (K, \omega_{sp})
		&=&
			- \frac{\epsilon\omega/c^2}{\lvert \gamma_1 \rvert}
				\left( 1 + \frac{\lvert \gamma_1 \rvert ^2}{\omega ^2/c^2} \right)
			- \frac{\epsilon \omega/c^2}{2\lvert \gamma_2 \rvert}
			+ \left(
				\lvert \gamma_1 \rvert \frac{1}{\omega}
				- \frac{\omega /c^2 }{2\lvert \gamma_2 \rvert}
			\right) \frac{\mathrm{d} \omega \epsilon}{\mathrm{d} \omega }
\end{eqnarray}
Using Eq.~\eqref{reldisp} and the definition of $\gamma_j$, we can simplify the terms
inside the parentheses and write
\begin{eqnarray}
	R ^{-1} (K, \omega_{sp})
		&=&
			\frac{\epsilon\omega/c^2}{\lvert \gamma_1 \rvert}
				\frac{\epsilon}{\lvert 1+ \epsilon \rvert}
			- \frac{\epsilon \omega/c^2}{2\lvert \gamma_2 \rvert}
			- \frac{\omega /c^2 }{2\lvert \gamma_2 \rvert}
				\frac{1-\epsilon}{1+\epsilon}
			 \frac{\mathrm{d} \omega \epsilon}{\mathrm{d} \omega }
\end{eqnarray}
Multiplying this expression by $\frac{\omega}{K^2} = \frac{c^2}{\omega}\frac{\epsilon +1}{\epsilon}$,
we get
\begin{eqnarray}
	\frac{\omega}{K^2} R ^{-1} (K, \omega_{sp})
		&=&
			\frac{-\epsilon}{\lvert \gamma_1 \rvert}
			- \frac{\epsilon +1}{2\lvert \gamma_2 \rvert}
			- \frac{1}{2\lvert \gamma_2 \rvert}
				\frac{1-\epsilon}{\epsilon}
			 \frac{\mathrm{d} \omega \epsilon}{\mathrm{d} \omega }
\end{eqnarray}
Comparing this expression to Eq.~\eqref{NormalizationL}, we can write
\begin{eqnarray}
	\frac{\omega}{K^2} R ^{-1} (K, \omega_{sp})
		&=& 2L(\omega)
\end{eqnarray}
which gives Eq.~\eqref{RL}.
\end{widetext}

\begin{acknowledgments}
\corr{The authors acknowledge C'nano Ile-de-France and
the French Agence Nationale de la Recherche (ANR) for financial support through the project LAPSUS.
Alexandre Archambault acknowledges financial support from the French Ministry of Defense
 through a grant from the Direction G\'en\'erale de l'Armement (DGA).}
\end{acknowledgments}


\begin{thebibliography}{99}


%1
\bibitem{Loudon} R. Loudon, \textit{The Quantum Theory of Light}, 3rd ed. (Oxford university press, New York, 2000)

%2
\bibitem{Raether}
H. Raether, \textit{Surface Plasmons on Smooth and Rough Surfaces and on Gratings} (Springer-Verlag, Berlin, 1988).

%3
\bibitem{Powell} C.J. Powel, J.B. Swan, Phys. Rev.\textbf{115}, 869 (1959)

%4
\bibitem{Lukin} A.V. Akimov, A. Mukherjee, C.L. Yu, D.E. Chang, A.S. Zibrov, P.R. Hemmer, H. Park, M.D. Lukin, Nature (London) \textbf{450} 402 (2007)

%5
\bibitem{Wrachtrup} R. Kolesov, B. Grotz, G. Balasubramanian, R.J. Stohr, A.A.L. Nicolet, P.R. Hemmer, F. Jelezko and J. Wrachtrup, Nature Phys. \textbf{5} 470-474 (2009)

%6
%7
\bibitem{Smuk2006125}
A.Y. Smuk and N.M. Lawandy, Appl. Phys. B \textbf{84}, 125-129 (2006)

%8
\bibitem{Li2005115409}
K. Li, X. Li, M.I. Stockman and D.J. Bergman, Phys. Rev. B \textbf{71}, 115409 (2005)

%9
\bibitem{Bergman2003027402}
D. J. Bergman and M. I. Stockman, Phys. Rev. Lett. \textbf{90}, 027402 (2003)

%...
\bibitem{ZheludevLasingSpaser}
N. I. Zheludev, S. L. Prosvirnin, N. Papasimakis and V. A. Fedotov, Nature Photon. 2, 351-354 (2008).


%10
\bibitem{Noginov2007455}
M.A. Noginov, M. Zhu, G. Bahoura, J. Adegoke, C. Small, B.A. Ritzo, V.P. Drachev and V.M. Shalaev, Appl. Phys. B \textbf{86}, 455-460 (2007)

%11
\bibitem{Seidel2005177401}
J. Seidel, S. Grafstrom and L. Eng, Phys. Rev. Lett. \textbf{94}, 177401 (2005)

%12
\bibitem{Ambati} M. Ambati, S.H. Nam, E. Ulin-Avila, D.A. Genov, G. Bartal and X. Zhang, Nano Lett. \textbf{8}, 3998-4001 (2008)

%13
\bibitem{NoginovPRL} M.A. Noginov, G. Zhu, M. Mayy, B.A. Ritzo, N. Noginova and V.A. Podolskiy, Phys. Rev. Lett. \textbf{101}, 226806 (2008) 

%14
\bibitem{Berini} I. De Leon and P. Berini, Phys. Rev. B \textbf{78}, 161401 (2008)

%15
\bibitem{Oulton} M. Ambati, D.A. Genov, R. Oulton and X. Zhang, IEEE J. of selected topics in quantum electronics \textbf{14}, 1395-1403 (2008)

%16
\bibitem{Stockman} K.F. MacDonald, Z.L. Samson, M.I. Stockman, N.I. Zheludev, Nature Photon. \textbf{3}, 55-58 (2009)

%17
\bibitem{Brongersma} L. Cao and M.L. Brongersma, Nature Photon. \textbf{3}, 12-13 (2009)

%18
\bibitem{Protsenko2005063812}
I.E. Protsenko, A.V. Uskov, O.A. Zaimidoroga, V.N. Samoilov and E.P. O'Reilly, Phys. Rev. A \textbf{71}, 063812 (2005)

%19
\bibitem{Bergman2004409}
D.J. Bergman and M.I. Stockman, Laser Phys. \textbf{14}, 409-411 (2004)

%
\bibitem{NoginovLaser}
M. A. Noginov, G. Zhu, A. M. Belgrave, R. Bakker, V. M. Shalaev, E. E. Narimanov, S. Stout, E. Herz,
T. Suteewong, and U. Wiesner, Nature \textbf{460}, 1110-1112 (2009).


%
\bibitem{OultonLaser}
Rupert F. Oulton, Volker J. Sorger, Thomas Zentgraf, Ren-Min Ma, Christopher Gladden, Lun Dai,
Guy Bartal, and Xiang Zhang, Nature \textbf{461}, 629-632 (2009).


%20
\bibitem{Barnes2002} W.L. Barnes, G. Bjork, J.M. Gerard, P. Jonsson, J.A.E. Wasey, P.T. Worthing and V. Zwiller, Eur. Phys. J. D \textbf{18}, 197-210 (2002)

%21
\bibitem{fasel} S. Fasel, M. Halder, N. Gisin and H. Zbinden, New J. of Phys. \textbf{8}, 13 (2006)

%22
\bibitem{Lukin2} D.E. Chang, A.S. Sorensen, P.R. Hemmer, M.D. Lukin, Phys. Rev. Lett. \textbf{97} 053002 (2006)

%23
\bibitem{Vuckovic} Y. Gong, J. Vuckovic, Appl. Phys. Lett. \textbf{90} 033113 (2007)

%24
\bibitem{Altewischer} E. Altewischer, M.P. van Exter, J.P. Woerdman, Nature (London) \textbf{418} 304 (2002)

%25
\bibitem{Moreno} E. Moreno, F.J. Garcia-Vidal, D. Erni, J. Ignacio Cirac, L. Martin-Moreno, Phys. Rev. Lett. \textbf{92} 236801 (2004)

%26
\bibitem{Fasel} S. Fasel, F. Robin, E. Moreno, D. Erni, N. Gisin, H. Zbinden Phys. Rev. Lett. \textbf{94} 110501 (2005)

%27
\bibitem{Zayats} M.S. Tame, C. Lee, J. Lee, D. Ballester, M. Paternostro, A.V. Zayats, M.S. Kim, Phys. Rev. Lett. \textbf{101} 190504 (2008)

%28
\bibitem{Elson19714129}
J.M. Elson and R.H. Ritchie, Phys. Rev. B \textbf{4}, 4129-4138 (1971)

%29
\bibitem{Gruner19961818}
T. Gruner and D.G. Welsch, Phys. Rev. A \textbf{53}, 1818-1829 (1996)

%30
\bibitem{Babiker19932236} M. Babiker, N.C. Constantinou and B.K. Ridley, Phys. Rev. B \textbf{48}, 2236-2243 ((1993)

%31
\bibitem{Stallinga2006026606}
S. Stallinga, Phys. Rev. E \textbf{73}, 026606 (2006)

%32
\bibitem{Landau} L. Landau and E. Lifchitz, \textit{Electrodynamics of continuous media}, 2nd ed. (Pergamon, Oxford, 1984)

%33
\bibitem{PRBAlex} A. Archambault, T. Teperik, F. Marquier, J.J. Greffet, Phys. Rev. B \textbf{79}, 195414 (2009)

%34
\bibitem{Palik} E. D. Palik, \textit{Handbook of Optical Constants of Solids} 
(Academic Press Inc., San Diego, 1985)

%35
\bibitem{Jackson} J.D. Jackson, \textit{Classical Electrodynamics}, 3rd ed. (John Wiley \& Sons, New York, 1999)

%36
\bibitem{Brillouin} L. Brillouin, \textit{Wave propagation and group velocity} (Academic Press, 1960)

%37
\bibitem{Purcell} E. M. Purcell, Phys. Rev. \textbf{69}, 681 (1946).

%38
\bibitem{Ford1984195}
G.W. Ford and W.H. Weber, Phys. Rep. \textbf{113}, 195-287 (1984)

%39
\bibitem{Novotny}
L. Novotny and B. Hecht, \textit{Principles of nano-optics} (Cambridge university press, New York, 2006)

\end{thebibliography}
\end{document}